\documentclass[twocolumn]{aastex701}

\usepackage{graphicx}
\usepackage{xspace}
\usepackage{amsmath}
\usepackage{xcolor}
\usepackage{subcaption}         
\usepackage{placeins}           

\newcommand{\soft}[1]{\texttt{#1}}
\newcommand{\unit}[1]{\ensuremath{\mathrm{\,#1}}\xspace}

\newcommand{\masyr}{\unit{mas\,yr^{-1}}}
\newcommand{\kpc}{\unit{kpc}\xspace}
\newcommand{\gaia}{{\it Gaia}\xspace}

\newcommand{\feh}{\unit{[Fe/H]}}

\newcommand{\vect}[1]{\boldsymbol{#1}}
\usepackage{csquotes}

\shorttitle{Mapping the Stellar Magellanic Stream with Distant K Giants}
\shortauthors{Cavieres et al.}

\begin{document}

   \title{The Distant Milky Way Halo from the Southern Hemisphere: Spatially broad and closer stellar debris associated to the Magellanic Stream\footnote{Based on observations collected at the European Southern Observatory under programme ID 115.28GR, and at Las Campanas Observatory, Carnegie Institution for Science, under Chilean TAC programs CN2025A-28 and CN2025B-95.}}


%





\author[0009-0002-2978-8383]{Manuel Cavieres}
\affiliation{Leiden Observatory, Leiden University, P.O. Box 9513, 2300 RA Leiden, the Netherlands}
\email[show]{cavierescarrera@strw.leidenuniv.nl}
\correspondingauthor{Manuel Cavieres}

\author[0009-0006-1167-9233]{Jes\'us Torres}
\affiliation{Instituto de Astrof\'{i}sica, Pontificia Universidad Cat\'{o}lica de Chile, Av. Vicu\~{n}a Mackenna 4860, 782-0436 Macul, Santiago, Chile}
\email{TODO} 

\author[0000-0002-4777-9934]{Camila Navarrete}
\affiliation{Departamento de F\'isica, Universidad T\'ecnica Federico Santa Mar\'ia, Vicu\~{n}a Mackenna 3939, San Joaqu\'in, Santiago de Chile, Chile}
\email{camila.navarrete@usm.cl} 

\author[0009-0009-0158-585X]{Himansh Rathore}
\affiliation{Department of Astronomy and Steward Observatory, University of Arizona, 933 N. Cherry Ave, Tucson, AZ 85721, USA}
\email{himansh@arizona.edu} 

\author[0000-0003-2481-4546]{Julio Chanam\'e}
\affiliation{Instituto de Astrof\'{i}sica, Pontificia Universidad Cat\'{o}lica de Chile, Av. Vicu\~{n}a Mackenna 4860, 782-0436 Macul, Santiago, Chile}
\email{TODO} 

\author[0000-0003-0715-2173]{Gurtina Besla}
\affiliation{Department of Astronomy and Steward Observatory, University of Arizona, 933 N. Cherry Ave, Tucson, AZ 85721, USA}
\email{TODO} 

\author[0000-0001-7107-1744]{Nicol\'as Garavito-Camargo}
\affiliation{Instituto de Astrof\'{i}sica, Pontificia Universidad Cat\'{o}lica de Chile, Av. Vicu\~{n}a Mackenna 4860, 782-0436 Macul, Santiago, Chile}
\email{TODO} 

\author[0000-0003-4341-6172]{A. Katherina Vivas}
\affiliation{Cerro Tololo Inter-American Observatory/NSF’s NOIRLab, Casilla 603, La Serena, Chile}
\email{TODO} 

\begin{abstract}
The Magellanic Stream (MS) is a gaseous structure extending over $150^\circ$ across the sky, likely formed through interactions between the Large and Small Magellanic Clouds (LMC and SMC) as they fell into the Milky Way (MW). Its predicted stellar counterpart would provide a collisionless record of these interactions, but few candidate members have been identified. To expand this sample, we constructed a catalog of 87,763 distant K-giant candidates using DELVE DR2 and VHS DR5 photometry and Gaia DR3 astrometry, and analyzed VLT and Magellan spectra of 51 targets along the trailing MS. Combining radial velocities, spectrophotometric distances, and proper motions, we identify five new MS candidates at 58–95 kpc. Their metallicities are consistent with the outskirts of the SMC, while orbit integrations in a time-dependent MW–LMC–SMC potential support a possible Magellanic origin. Combined with previously identified candidates, these stars span a substantially broader range of Magellanic latitude than predicted by existing models, suggesting an extended debris field rather than a narrow stream. The candidates also lie predominantly closer than the simulated debris. More recent, less eccentric SMC orbits may help explain this distance discrepancy, although new simulations are required to test their implications for the stellar debris, along with larger spectroscopic samples.
\end{abstract}

   \keywords{\uat{Milky Way stellar halo}{1060} --- \uat{Magellanic Stream}{991} --- \uat{Magellanic Clouds}{990} --- \uat{Galaxy interactions}{600}}

\section{Introduction}
\label{sec:introduction}

The LMC and SMC are the nearest example of an ongoing interaction between massive satellite galaxies and a Milky Way (MW)-mass host \citep{Besla2012,DOnghia16}. Their proximity makes the Clouds
uniquely valuable laboratories for studying satellite accretion, dwarf--dwarf interactions and gas removal
\citep{Besla2007,Besla2012}, as well as the dynamical response of a galaxy halo to a massive perturber \citep[e.g.,][]{GaravitoCamargo2019, Petersen20, Petersen2021, Foote2023, Rathore_2026}.

The proper motions of the Clouds indicate that they must have interacted recently \citep{Kallivayalil2013, Zivick2018, Patel2020}. During this galaxy--galaxy interaction, tidal theory demands the formation of tidal bridges and tails \citep{Toomre1972}. A gaseous bridge between the Clouds and a stream trailing them are both known to exist \citep{Mathewson1974, Putman2003, Nidever2010, mishra2025}. Every simulation of the Clouds' interaction history predicts the formation of a less prominent stellar tidal tail (the stellar MS), in addition to a gaseous tidal tail \citep[e.g.,][]{Lin1995, Gardiner1996, Ruzica2010, DiazBekki2012, Besla2013, luccini2021}.

Spectroscopic follow-up of distant halo samples have revealed stars potentially associated to the stellar MS \citep{Zaritsky2020}, although their kinematics can also be associated to the GES merger \citep{Chandra2023}. More recently, through spectroscopic followup of K giants selected from WISE \citep{wise_survey} and \gaia \citep{gaiadr3}, \citet{Chandra2023} were able to identify 13 high-confidence members of the stellar counterpart to the gaseous trailing arm of the Stream (stellar MS). They combined spectroscopic distances, radial velocities, and metallicities with \gaia\ astrometry to obtain chemo-dynamical constraints. We refer to these thirteen stars as the Chandra sample (labeled Chandra+23 in the figures). This discovery suggests that a stellar MS may now be detected, but the existing spectroscopic samples remain small. A larger sample is therefore needed to constrain the kinematic structure of the debris, its extent across the sky, and its distance.

 If stars can be confidently associated with the MS, their positions, distances, radial velocities, proper motions, and ages can provide collisionless constraints on the timing, geometry, and strength of the LMC--SMC interaction. The gas alone cannot provide these constraints because gas and stars respond to different physical processes \citep[for a review, see][]{DOnghia16}. Gas can be displaced by ram pressure \citep{Salem2015}, cooling and feedback \citep{Nidever2008}, and interaction with the circumgalactic medium \citep{Fox2014, Lucchini2020}, whereas stars are essentially collisionless tracers of the gravitational field \citep{DiazBekki2012, Besla2013} and therefore provide critical constraints on the orbital geometry of the system.

In this paper, we present the results of a new search for stellar members of the MS. We selected distant K-giants using optical photometry from the DECam Local Volume Exploration Survey \citep[DELVE,][]{DrlicaWagner2022} and near-infrared photometry from the VISTA Hemisphere Survey \citep[VHS,][]{McMahon2013}. This new optical--near-infrared selection reaches fainter magnitudes than earlier Gaia--WISE selections and enables the identification of candidate giants at heliocentric distances comparable to those expected for Magellanic debris. 

We obtained medium-resolution spectra of a subsample of these K-giants to measure radial velocities and metallicities. These quantities provide the key tests of a Magellanic origin. Stars stripped from the SMC should occupy the appropriate region of phase space, follow a coherent velocity trend along the Stream, and have metallicities consistent with Magellanic stellar populations. More importantly, confirmed stellar debris can be used as a dynamical tracer of the SMC's past orbit. Unlike the gas, these old stars retain a largely collisionless memory of the stripping event, making them especially useful for testing models of the latest LMC--SMC interaction.

The paper is organized as follows. Section~\ref{sec:data} describes the photometric and astrometric data and the quality cuts applied. Section~\ref{sec:selection} presents the selection of distant halo K giants, and Section~\ref{sec:xshooter} describes their spectroscopic follow-up. Section~\ref{sec:results} examines the stars' association with the Clouds using their 3D positions and motions. We summarize our results in Section~\ref{sec:conclusions}.
\section{Survey data}
\label{sec:data}
In this section, we describe the photometric and astrometric surveys used for this study. 
The photometric data come mainly from the second data release of DELVE \citep{delve}, which was a survey program on the 4-m Blanco Telescope at the Cerro Tololo Inter-American Observatory in Chile. DELVE covers $\sim 15,000$ $\deg^2$ of the Southern sky at high Galactic latitude ($|b| > 10 \deg$). It provides photometry in the $g$, $r$, $i$, and $z$ bands to a $5\sigma$ depth of $\sim 23.5$ mag.

We cross-matched the second data release of the DELVE catalog with the 5th data release of VHS \citep{vhs}, which provides photometric coverage across the Southern celestial hemisphere in the $J$ and $K_{\rm S}$ near-infrared bands using the VIRCAM instrument, formerly installed on the VISTA telescope. These combined catalogs encompass most of the Southern celestial hemisphere at $J < 17$ mag, with the necessary photometric bands to select distant halo tracers such as the K giants used in this study. To remove clear foreground objects using \gaia\ parallaxes, we also cross-matched this combined catalog with the third data release of the \gaia mission \citep{gaiadr3}.

To ensure reliable measurements, the DELVE query required point-source classifications in the $g$, $i$, and $z$ bands and zero photometric flags in the $g$, $r$, $i$, and $z$ bands. The final catalog additionally required $J<17$. Where overlapping VHS tiles provide more than one detection of the same \gaia source, we kept a single entry, preferring the detection flagged as primary in VHS (\texttt{PRIMARY\_SOURCE}$\,=1$). These selections are listed in Table~\ref{tab:quality}.

\begin{table}[h]
\centering
\caption{Quality selections used to define the final 87\,763-object catalog.} \label{tab:quality}
\begin{tabular}{ll}
    \hline
    Survey & Selection criteria \\
    \hline
    DELVE & $\texttt{extended\_class\_g} = 0$ \\
    DELVE & $\texttt{extended\_class\_i} = 0$ \\
    DELVE & $\texttt{extended\_class\_z} = 0$ \\
    DELVE & $\texttt{flags\_g} = 0$ \\
    DELVE & $\texttt{flags\_r} = 0$ \\
    DELVE & $\texttt{flags\_i} = 0$ \\
    DELVE & $\texttt{flags\_z} = 0$ \\
    VHS & $\texttt{JAPERMAG4} < 17$ \\
    VHS & one detection per \gaia~DR3 source \\
    \hline
\end{tabular}
\end{table}

The observed DELVE and VHS magnitudes were used in the color--color and color--magnitude diagrams, and preliminary photometric-distance selections.

\section{Halo K-giant selection}
\label{sec:selection}

\begin{figure}
    \centering
    \includegraphics[width=\linewidth]{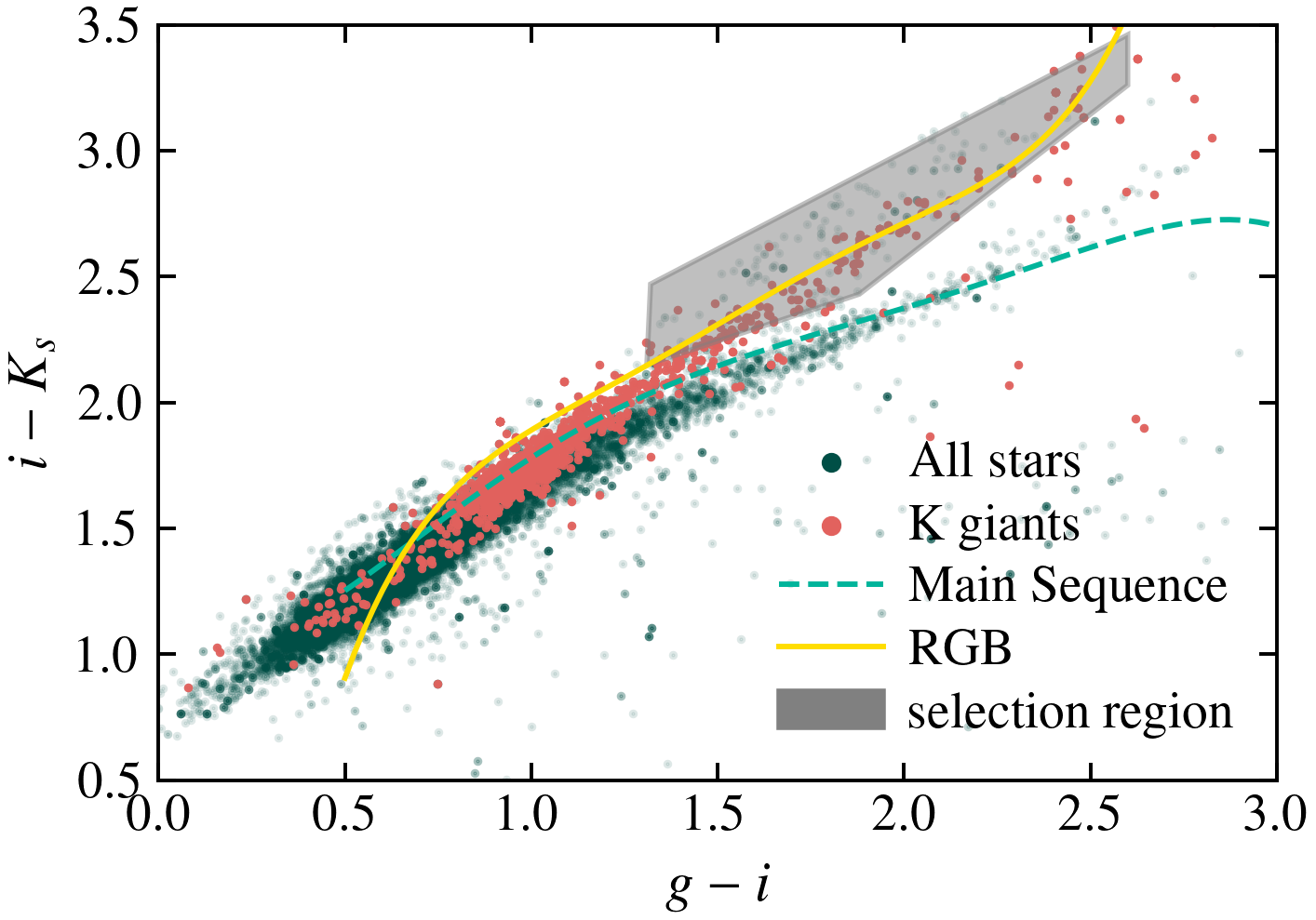}
    \caption{\textit{giKs} color--color diagram showing the selection region near the tip of the RGB as a gray-shaded area. Spectroscopically selected K giants from \citet{kgiants_zhang} are shown as red points. A PARSEC isochrone for a 10 Gyr, $\mathrm{[M/H]}=-0.9$ population is shown in green for the main sequence and yellow for the RGB.}
    \label{fig:giKs}
\end{figure}

We selected K giants based on their position in the \textit{giKs} diagram, where the tip of the red giant branch (RGB) separates from the main stellar locus, as shown in Figure~\ref{fig:giKs}. The \textit{giKs} diagram was first presented by \citet{wake} as an effective tool to separate stellar sources from unresolved galaxies. It was developed based on ($u-i$), ($i-Ks$) color-color diagrams, extensively used to separate stars and galaxies at large distances \citep[see][]{Munoz14, 2018ApJ...860....4O}, and the identification of K giant stars based on mid-infrared WISE photometry in \cite{Conroy2021}. In this plane, metal-poor RGB stars are offset towards redder $i-K_s$ colors at fixed $g-i$ relative to main-sequence stars, which allows K giants to be selected using only $g$, $i$, and $K_s$ bands \citep{wake}. The K giants were selected to lie in the shaded gray region shown in Figure~\ref{fig:giKs}, corresponding to the interior of the polygon defined by:

\begin{equation}
\mathcal{P}_{(g-i,\; i-K_s)} =
\left\{
\begin{aligned}
&(1.31,\;2.15),\\
&(1.88,\;2.43),\\
&(2.60,\;3.26),\\
&(2.60,\;3.46),\\
&(1.32,\;2.47)
\end{aligned}
\right\}.
\end{equation}

This selection, motivated by the position of the RGB of a PARSEC 1.2S \citep{parsec} isochrone in this diagram (age = 10 Gyr, $\mathrm{[M/H]}=-0.9$), allows stars at the tip of the RGB to be separated from dwarfs through their infrared excess, while limiting contamination from unresolved background galaxies, which dominate the counts at faint magnitudes.

The final DELVE-VHS-\gaia catalog contains 87\,763 K-giant candidates distributed over an effective sky area of approximately $1.4\times10^{4}\,\mathrm{deg}^{2}$, corresponding to a mean surface density of $6.3$ stars $\mathrm{deg}^{-2}$. The footprint spans all Galactic longitudes and extends from $b=-86.8^\circ$ to $+58.4^\circ$, with 93.9\% of the sources located at $|b|>10^\circ$. The catalog covers $15.0<g<22.4$ mag in DELVE PSF magnitudes, has a median magnitude of $g=19.3$, and contains 23\,989 sources satisfying the foreground-rejection criterion $\varpi<0.1$ mas, corresponding to $\approx 10 \,\mathrm{kpc}$. The full catalog, including the photometric distances of Section~\ref{sec:photometric_distances}, will be made publicly available on Zenodo\footnote{\url{https://doi.org/10.5281/zenodo.24083695}}.



\subsection{Photometric distances}
\label{sec:photometric_distances}

We estimated a preliminary photometric distance for each K-giant candidate by treating it as an RGB star. We linearly interpolated the RGB of a single 10~Gyr, $\mathrm{[Fe/H]}=-1.5$ PARSEC~v1.2S isochrone \citep{parsec} in the $(g-i,M_g)$ plane and converted $g-M_g$ to a distance modulus. These point estimates, computed from the observed photometry, were used for target selection. Stars whose $g-i$ color lies outside the color range of the isochrone RGB have no photometric distance; this applies to 4\,267 of the 23\,989 sources with $\varpi<0.1$~mas.




\begin{figure}
    \centering
    \includegraphics[width=\linewidth]{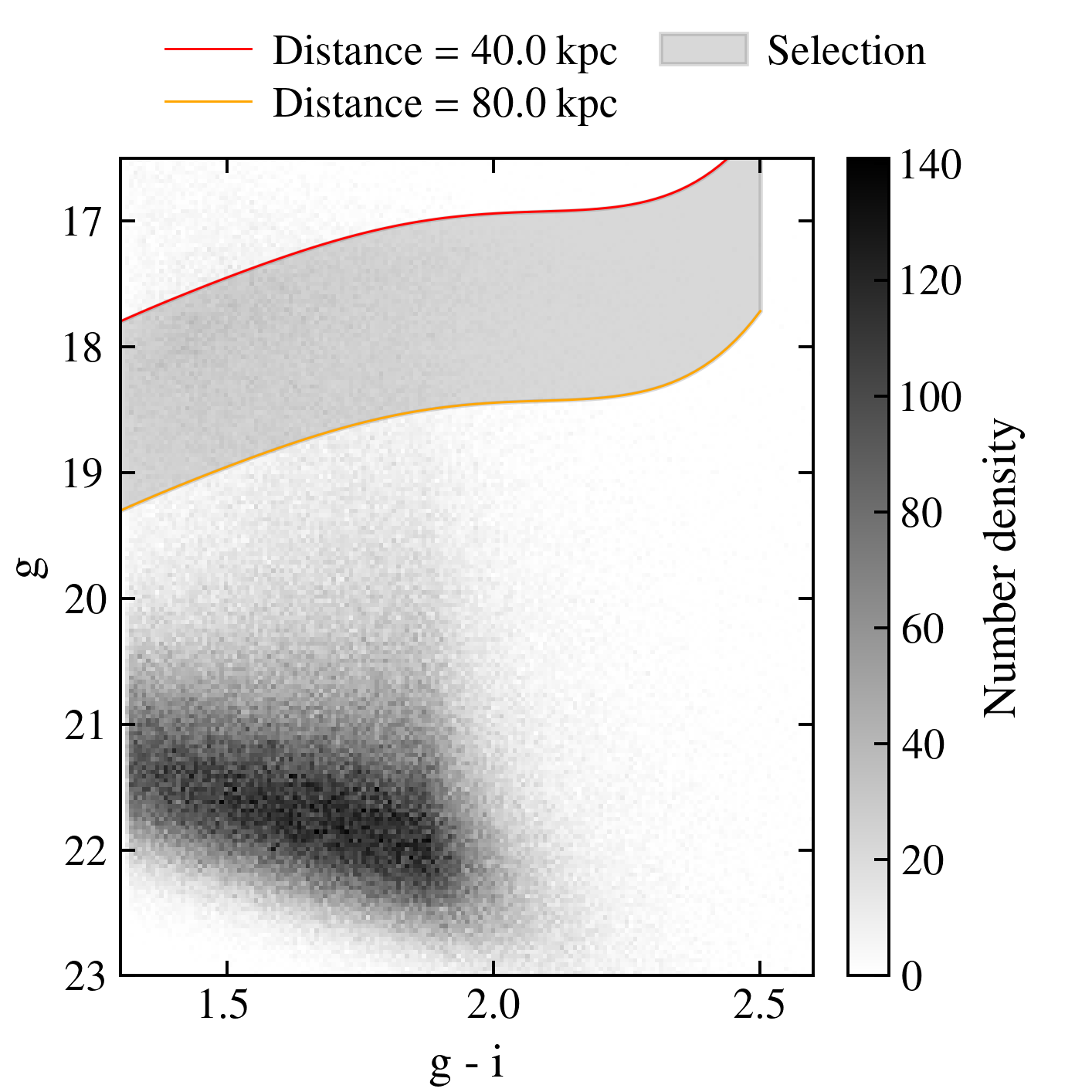}
    \caption{Isochrone distance-based selection of K-giant stars for the spectroscopic follow-up. The two ridgelines correspond to a 10 Gyr, $\mathrm{[M/H]}=-0.9$ PARSEC~v1.2S population placed at 40 and 80~kpc. Selecting in this distance regime retains distant objects without incorporating contaminants at fainter magnitudes.}
    \label{fig:isochrone_selection_observations}
\end{figure}


\section{Spectroscopic follow-up}
In this section we describe how we selected targets from the K-giant catalog for spectroscopic follow-up of the stellar MS. We present the observations and the methods used to obtain stellar parameters, radial velocities, metallicities and spectro-photometric distances.

\label{sec:xshooter}
\subsection{Candidate selection for X-shooter follow-up}

Follow-up targets were required to be distant, to lie along the trailing arm of
the Stream, and to have a \gaia~DR3 proper motion compatible with that of
Magellanic debris. The criteria are collected in
Table~\ref{tab:kgiant_selection} and described below in the order in which they
were applied.

Foreground stars were removed with a parallax cut, $\varpi<0.1$~mas, and the
sample was restricted to the Southern Galactic hemisphere ($b<0^\circ$), where
the trailing arm is found. Distant giants were then isolated directly in the
color--magnitude diagram. We
fitted a fifth-order polynomial to the RGB of a 10~Gyr,
$\mathrm{[M/H]}=-0.9$ PARSEC~v1.2S isochrone \citep{parsec} in the $(g-i,\,g)$
plane, shifted the resulting RGB to heliocentric distances of 40 and
80~kpc, and kept the 11\,993 stars falling between the two shifted ridgelines.
This band, shown in Figure~\ref{fig:isochrone_selection_observations}, brackets
the distance range in which Magellanic debris is expected along the trailing
arm, while its faint edge keeps the selection above the magnitude regime where
unresolved background galaxies dominate the counts.

Sky positions and proper motions were then transformed into the Magellanic
Stream coordinate system of \citet{Nidever2010}, and a deliberately generous
window was applied around the trailing arm position, $-125^\circ<L_{\rm MS}<-25^\circ$
and $-50^\circ<B_{\rm MS}<20^\circ$. The longitude range excludes the bodies of
the Clouds themselves while covering the full extent of the trailing arm within
the survey footprint, and the latitude range was chosen to be wide and asymmetric
about the gaseous Stream, so as not to assume that the stellar debris
follows strictly the spatial distribution of H\,\textsc{i}. This window contains 150 entries in the catalog, 
corresponding to 126 distinct \gaia sources; the 24 sources appearing twice, as
a consequence of repeated DELVE or VHS detections of the same star in
overlapping tiles, were discarded altogether to avoid any ambiguity in the
target coordinates, leaving 102 candidates.

Two kinematic cuts were then imposed on the proper motions projected onto the
Stream frame: $0<\mu_L<3$~mas~yr$^{-1}$, which requires motion along the Stream
in the sense of the trailing arm, and $|\mu_B|<0.5$~mas~yr$^{-1}$, which
requires that motion to be confined to the Stream plane. These cuts leave 97 and 85
candidates, respectively. Finally, we imposed a distance--longitude window
following the distance gradient expected along the trailing arm, keeping stars
with $d>50-0.3\,(L_{\rm MS}+25^\circ)$~kpc and
$d<80-1.2\,(L_{\rm MS}+25^\circ)$~kpc, where $d$ is the preliminary photometric distance inferred from the 10~Gyr, $\mathrm{[Fe/H]}=-1.5$ RGB isochrone described in Section~\ref{sec:selection}, we combine both distance estimates as one represents a general halo star age and metallicity, while the other assumes an SMC halo-like star. These two straight lines in the $(L_{\rm MS},d)$
plane generously bracket the distance gradient predicted for the trailing arm
by the \citet[][hereafter B+12]{Besla2012} models, widening from $50$--$80$~kpc at
$L_{\rm MS}=-25^\circ$ to $80$--$200$~kpc at $L_{\rm MS}=-125^\circ$. This last
cut yields the final X-shooter target list of 51 stars, spanning
$-101.6^\circ<L_{\rm MS}<-28.6^\circ$, $-43.1^\circ<B_{\rm MS}<16.7^\circ$,
$57<d<112$~kpc and $17.2<g<18.8$. The 12.8~h awarded to the program allowed us to observe 33 of them. These stars span $17.2<g<18.5$ and $57<d<99$~kpc and are not a strictly magnitude-limited subset of the target list; they constitute the more targeted VLT search discussed in the rest of the paper.

\begin{table}
\caption{Selection criteria applied to the K-giant catalog to define the
X-shooter target list, in the order in which they were applied. Here $d$ is
the preliminary isochrone-based photometric distance described in
Section~\ref{sec:photometric_distances}, obtained by interpolating a 10~Gyr,
$\mathrm{[Fe/H]}=-1.5$ PARSEC isochrone in the $(g-i,\,M_g)$ plane. The RGB ridgelines instead define the color--magnitude band using the $\mathrm{[M/H]}=-0.9$ isochrone described in the text.}
\label{tab:kgiant_selection}
\label{table:cuts}
\centering
\begin{tabular}{l}
\hline\hline
Criterion \\
\hline
$\varpi < 0.1~\mathrm{mas}$ \\

$b < 0^\circ$ \\

Between the 40 and 80~kpc RGB ridgelines \\

$-125^\circ < L_{\rm MS} < -25^\circ$ \\

$-50^\circ < B_{\rm MS} < 20^\circ$ \\


$0 < \mu_L < 3~\mathrm{mas~yr^{-1}}$ \\

$-0.5 < \mu_B < 0.5~\mathrm{mas~yr^{-1}}$ \\

$d > 50 - 0.3\left(L_{\rm MS}+25^\circ\right)~\mathrm{kpc}$ \\

$d < 80 - 1.2\left(L_{\rm MS}+25^\circ\right)~\mathrm{kpc}$ \\

\hline
\end{tabular}
\tablecomments{
$L_{\rm MS}$ and $B_{\rm MS}$ are
MS coordinates \citep{Nidever2010}, $\mu_L$ and $\mu_B$ the
corresponding proper-motion components.
}
\end{table}

\subsection{X-shooter observations}
We obtained medium-resolution spectra with the X-shooter instrument \citep{Vernet2011}, mounted on the Very Large Telescope. Échelle long-slit spectra were acquired for 33 stars over 12.8 h in program 115.28GR.001 (PI J. Chanamé), observed in Service Mode from July to September 2025. We used slit widths of 0\farcs8, 0\farcs7, and 0\farcs6 for the UVB, VIS, and NIR arms, respectively, yielding spectral resolutions of $\sim$6\,700, 11\,400 and 8\,100 in the respective arms. Exposure times ranged from 600 to 900 s, yielding a mean S/N of $\sim$10, $\sim$23, and $\sim$27 for UVB, VIS, and NIR spectra.

The reduced, order-merged, and extracted 1D spectra were retrieved from the X-shooter ESO Phase 3 Data Products. The raw data were reduced with the ESO pipeline \texttt{xshoo-2.3.12}, which performs bias (UVB, VIS arms) and dark (NIR arm) correction, cosmic-ray removal, flat-field correction, sky subtraction, wavelength calibration, and flux calibration. No telluric correction was performed; instead, regions affected by telluric absorption were masked, as explained in Section~\ref{sec:stellar_parameters}.

\subsection{Candidate selection for the MagE follow-up}
\label{sec:mage_selection}

A second target list was built to extend the coverage of the spectroscopic sample further along the
trailing arm than the X-shooter selection reached. The criteria are collected in
Table~\ref{tab:mage_selection} and described below in the order in which they
were applied. Starting from the 20\,001
sources with $\varpi<0.1$~mas whose VHS detection is the primary one (\texttt{PRIMARY\_SOURCE}$\,=1$), a subset of the 23\,989 such sources in the final catalog (Section~\ref{sec:selection}), we
repeated the color--magnitude selection of Section~\ref{sec:xshooter} with the
RGB ridgeline placed at $40$ and $160$~kpc rather than $40$ and $80$~kpc,
aiming to encompass the larger distances expected at the far end of the
arm. A total of 15\,452 stars fall between the two ridgelines. Neither the hemisphere
cut nor the trailing-arm window of Section~\ref{sec:xshooter} was imposed at
this stage: we required only $-135^\circ<L_{\rm MS}<80^\circ$ and
$|B_{\rm MS}|<50^\circ$, leaving 10\,764 stars. The kinematic criteria were
likewise relaxed to $-1<\mu_L<3$~\masyr, which admits motion along the Stream
in either sense, and $|\mu_B|<0.5$~\masyr, leaving 10\,411 and 6\,870
candidates, respectively.

Restricting the sample to the trailing arm, $L_{\rm MS}<-25^\circ$, yields
$136$ targets, of which $27$ had already been observed with X-shooter and were
removed. The remaining $109$ candidates span
$-103.7^\circ<L_{\rm MS}<-25.2^\circ$ and $17.0<g<19.9$ mag; of these, 28 were observable during the allocated time (Section~\ref{sec:mage_observations}) and were observed; 18 yielded reliable radial velocities and atmospheric
parameters. The resulting MagE spectroscopic sample spans $-95.9^\circ<L_{\rm MS}<-25.2^\circ$,
$-24.1^\circ<B_{\rm MS}<23.7^\circ$, $17.0<g<19.6$ mag and $55<d<161$~\kpc.

\begin{table}
\caption{Selection criteria applied to the K-giant catalog to define the
MagE target list, in the order in which they were applied. The criteria
deliberately relax those of Table~\ref{tab:kgiant_selection} so as to reach
further along the trailing arm; see Section~\ref{sec:mage_selection}. The RGB ridgelines define the color--magnitude band using the same $\mathrm{[M/H]}=-0.9$ isochrone as in Table~\ref{tab:kgiant_selection}, shifted to 40 and 160~kpc.}
\label{tab:mage_selection}
\centering
\begin{tabular}{l}
\hline\hline
Criterion \\
\hline
$\varpi < 0.1~\mathrm{mas}$ \\

Between the 40 and 160~kpc RGB ridgelines \\

$-135^\circ < L_{\rm MS} < 80^\circ$ \\

$|B_{\rm MS}| < 50^\circ$ \\

$-1 < \mu_L < 3~\mathrm{mas~yr^{-1}}$ \\

$|\mu_B| < 0.5~\mathrm{mas~yr^{-1}}$ \\

$L_{\rm MS} < -25^\circ$ \\

\hline
\end{tabular}
\tablecomments{
$L_{\rm MS}$, $B_{\rm MS}$, $\mu_L$ and
$\mu_B$ are as in Table~\ref{tab:kgiant_selection}. The longitude, latitude and
proper-motion windows are wider than their X-shooter counterparts; in
particular, the $\mu_L$ range admits motion along the Stream in either sense.
}
\end{table}

\subsection{MagE observations}\label{sec:mage_observations}

We carried out this second spectroscopic follow-up with the Magellan Échelle Spectrograph \citep[MagE;][]{Marshall08} mounted on the 6.5~m Magellan Baade telescope at Las Campanas Observatory, as part of the program CN2025B-95 (PI J. Chanamé), over two observing nights (early November 2025 and early January 2026). We used a 0\farcs7 slit width, yielding a spectral resolution of $\sim$5\,850, with magnitude-dependent exposure times ranging from 15 to 40 minutes to achieve a minimum S/N of $\sim$15 at $5\,100 \,\AA$.

The data were reduced with the \soft{PypeIt} \texttt{1.17.3} pipeline \citep{Prochaska2020, PypeIt25}, which carries out bias and flat-field correction, wavelength calibration using paired ThAr arc exposures, and optimal extraction of each order. Following the standard \soft{PypeIt} procedure, the individual orders were flux-calibrated and merged into a continuous spectrum. The continuum was estimated using the procedure implemented in the \soft{iSpec} package \citep{BlancoCuaresma2014, BlancoCuaresma2019}, adapted to the MagE spectra by adjusting the continuum-fitting parameters, including the instrumental resolution and spline degree. A spline function was fitted to the observed spectrum, with strong absorption features automatically identified and excluded from the fit based on local flux variations. Finally, the normalized spectrum was obtained by dividing the observed flux by the fitted continuum.

\subsection{Stellar parameters}
\label{sec:stellar_parameters}

We derived stellar atmospheric parameters and radial velocities from the X-shooter VIS and MagE spectra using the \soft{rvspecfit} package \citep{rvspecfit}. We compared each spectrum with PHOENIX v2.0 synthetic stellar spectra \citep{Husser2013}, jointly fitting the effective temperature ($T_{\rm eff}$), surface gravity ($\log g$), metallicity (\feh), $\alpha$-element abundance ($[\alpha/{\rm Fe}]$), and radial velocity. A smooth multiplicative polynomial was included to account for large-scale differences in continuum shape between the observations and models, ensuring that the fit was driven primarily by the absorption-line information.

\begin{figure}
    \centering
    \includegraphics[width=\linewidth]{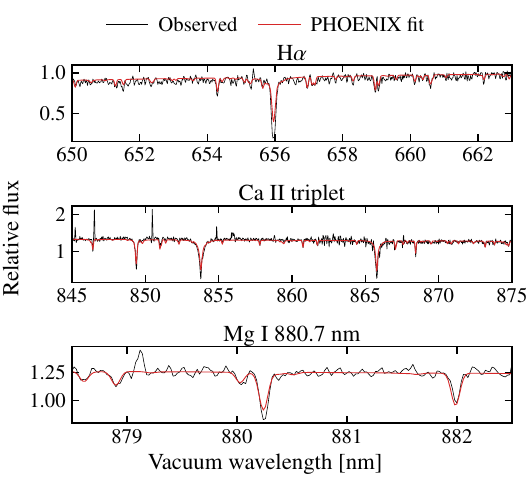}
    \caption{Flux-calibrated spectra of \gaia~DR3 6898149830259505408, zoomed in on the $H_\alpha$, Ca II triplet, and Mg I regions to show the quality of the \soft{rvspecfit} fit.}
    \label{fig:spec_zoom_example}
\end{figure}

Pixels with invalid uncertainties and the strong atmospheric O$_2$ A-band were excluded from the X-shooter fits. Regions affected by telluric absorption were identified and masked out using \soft{TelFit} \citep{telfit}. For the MagE data, we restricted our analysis to the spectral region between $5\,000$ and $6\,700\,\AA$, thereby avoiding regions of low S/N and less reliable wavelength calibration. Figure~\ref{fig:spec_zoom_example} highlights the agreement between the observed spectrum from XSHOOTER and its best-fitting PHOENIX model around H$\alpha$, the Ca~II triplet, and the Mg~I line at 880.7\, nm. The resulting $T_{\rm eff}$, $\log g$, and \feh measurements were subsequently used as spectroscopic constraints in the distance determination described below.

The covariance matrix returned by \soft{rvspecfit} is derived from
the local Hessian of the likelihood \citep{rvspecfit,Koposov2024}.
The corresponding uncertainties therefore quantify only the formal, local
statistical precision conditional on the adopted spectral model and do
not account for correlated-pixel noise, continuum-placement errors, or
imperfections and interpolation systematics in the PHOENIX templates
\citep{Li2019,Lancon2021,Koposov2024}. External validation of
\soft{rvspecfit} and X-shooter full-spectrum analyses indicates
atmospheric-parameter uncertainties of order $0.2$--$0.3$ dex, with the
largest discrepancies occurring for cool, low-gravity giants
\citep{Li2019,Arentsen2019,Koposov2024}. We therefore adopt
literature-informed systematic uncertainty floors of $0.30$ dex in
$\log g$, $0.20$ dex in \feh, and 100~K in $T_{\rm eff}$ combined in quadrature with the formal
uncertainties according to
$\sigma_{\rm tot}^{2}=\sigma_{\rm Hess}^{2}+\sigma_{\rm sys}^{2}$.

For the MagE spectra, we additionally compared our results with independent measurements for a small number of targets in common with other samples \citep[two stars with X-shooter analyses and one star, observed twice with MagE, from][]{Chandra2023}. Based on the scatter in \feh relative to these comparison stars, we adopt a larger systematic floor in \feh for the MagE sample ($0.30$ dex instead of $0.20$ dex).
For fits flagged with \texttt{bad\_hessian}, for which the local
covariance matrix could not be robustly estimated, we report the
adopted systematic uncertainty alone. The resulting atmospheric parameters are shown in Tables \ref{tab:high-likelihood-candidates} and \ref{tab:non-candidates}, for X-shooter and MagE data, respectively.

\subsection{Spectro-photometric distances}
\label{sec:dist}
We obtained spectro-photometric distances by fitting MESA Isochrones \& Stellar Tracks \citep[MIST,][]{mist1, mist2, mist3, mist4, mist5} to the spectroscopic constraints on effective temperature, surface gravity, and metallicity together with the photometry. We used the DELVE $g$, $r$, $i$, and $z$ bands; the VHS $J$ and $K_{\rm S}$ bands, with $H$ where available; and the \gaia\ G, BP, and RP bands. This gives nine photometric bands per star, or ten where $H$ is measured. The DELVE $u$ band is not used: no extinction coefficient is available for it, so it cannot be de-reddened consistently with the others. We performed the fits with the \soft{mistfit} package, which uses nested sampling and MIST models interpolated with \soft{minimint} \citep{minimint}. For details of the fitting, see Section~4.1.2 of \citet{hvs_desi}. The implementation is publicly available\footnote{\url{https://github.com/mncavieres/mistfit}}.

The resulting posterior from \soft{mistfit} includes mass, age, metallicity, distance, and extinction. We adopted a mass prior based on a Salpeter IMF \citep{Salpeter1955}, with allowed masses between 0.1 and $15\,M_{\odot}$; a flat prior for \feh within the range allowed by MIST; a flat prior on distance between 1 and 200~\kpc; and a uniform prior on extinction from zero to five times the extinction reported in \citet{sfd_og}, using the recalibration given in \citet{sfd_recal} and the corresponding relative extinction coefficients.

We further restricted the stellar age to 4--13~Gyr, consistent with the old populations expected in the Magellanic system, and sampled the isochrones in equivalent evolutionary phase rather than in mass and age, which covers the giant branch more uniformly. The \gaia\ parallax is evaluated in the likelihood rather than used to impose a prior, so that stars with low-significance or negative parallaxes are retained without truncation. An example posterior is shown in Figure~\ref{fig:mistfit}.


\section{Results}
\label{sec:results}

We use the spectroscopic sample of 51 stars described in Section~\ref{sec:xshooter} to identify stars consistent with the stellar MS and compare them with simulations and previous samples.

\begin{figure}
    \centering
    \includegraphics[width=\linewidth]{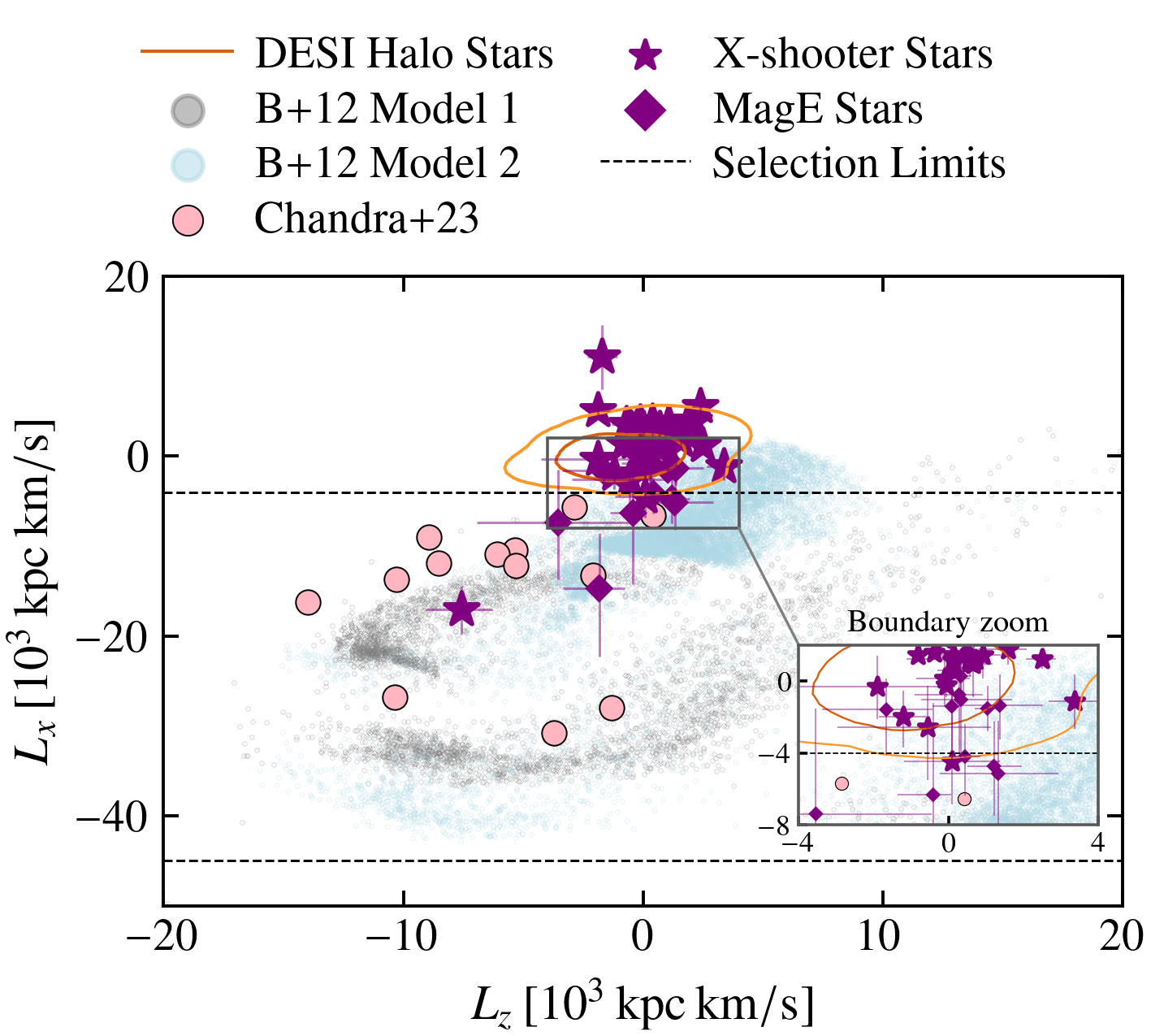}
    \caption{Angular momentum of the spectroscopic samples in the Galactocentric $x$ and
    $z$ directions. Purple stars mark the 33 X-shooter targets and purple diamonds the
    18 MagE targets.
    Pink circles show the Chandra sample \citep{Chandra2023}. Gray and
    light blue points are SMC star particles from the two B+12 models,
    with the Cloud main bodies removed. Orange contours trace the number density of
    DESI halo stars. The dashed lines bound the adopted central-value selection,
    $-45\times10^{3}\leq L_x\leq-4\times10^{3}\,\mathrm{kpc\,km\,s^{-1}}$.
    The inset enlarges the boxed region around $L_z=0$ near the upper selection boundary, using the same axis units as the main panel.
    Three MagE targets below the lower bound fall outside the plotted range.
    Eight stars pass this central-value band. }
    \label{fig:lz}
\end{figure}


Using the full six-dimensional phase-space information for these stars, we compute their orbital angular momenta to assess their association with the Magellanic Clouds, presented in figure \ref{fig:lz}, with error bars giving
the 16th and 84th percentiles of the angular-momentum posteriors from 1000
samples over the proper-motion, distance and radial-velocity uncertainties. Following \citet{Chandra2023}, we use this angular-momentum space as an initial means of distinguishing potential MS members from MW halo stars. Figure~\ref{fig:lz} shows the angular momentum about the Galactocentric $x$-axis, $L_{\rm x}$, as a function of the angular momentum about the $z$-axis, $L_{\rm z}$, for our targets, together with stellar particles from B+12 Models~1 and~2. To isolate the simulated stellar MS from the main bodies of the Clouds, we mask particles with Magellanic Stream longitudes $-20^\circ \leq L_{\rm MS} \leq 0^\circ$, retaining those with $L_{\rm MS}<-20^\circ$ or $L_{\rm MS}>0^\circ$. The corresponding angular-momentum distribution for a representative sample of MW halo stars is constructed using stars from the DESI Milky Way Survey \citep{Koposov2024} and is represented by the $1\sigma$ and $2\sigma$ contours in Figure~\ref{fig:lz}. The MW halo is concentrated primarily at $L_{\rm x}\gtrsim-4\times10^{3}\,\mathrm{kpc\,km\,s^{-1}}$. Eight stars pass our adopted central-value band, $-45\times10^{3}\leq L_{\rm x}\leq-4\times10^{3}\,\mathrm{kpc\,km\,s^{-1}}$: two of the 33 X-shooter targets and six of the 18 MagE targets. Three further MagE targets have $L_x$ below the lower bound and lie off the plotted panel.

\begin{figure}
    \centering
    \includegraphics[width=\linewidth]{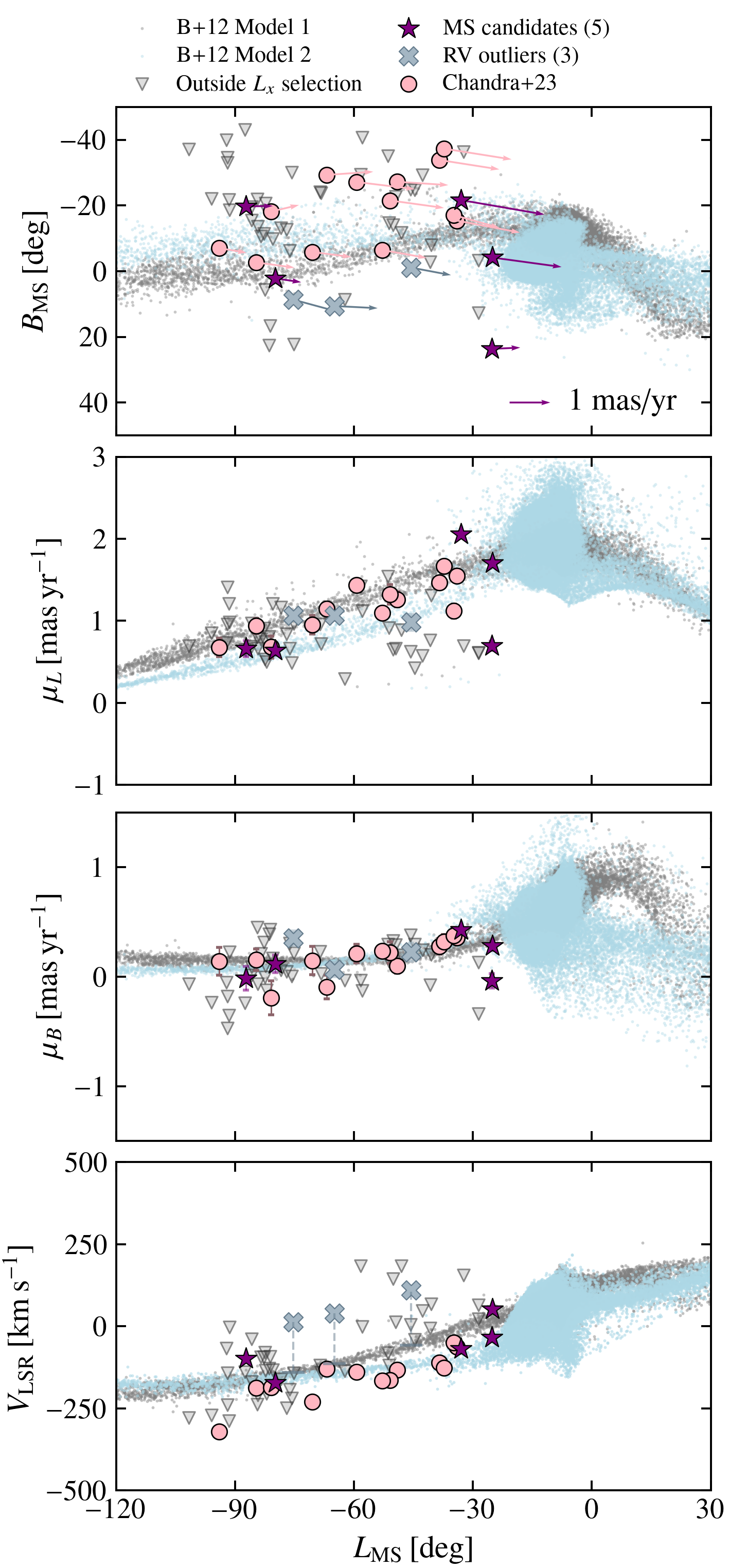}
    \caption{Magellanic latitude, proper motions, and LSR radial velocity versus Magellanic longitude. Purple stars are the five MS candidates, blue-gray crosses the three radial-velocity outliers, and gray triangles the other 43 targets. Pink circles show the Chandra sample \citep{Chandra2023}; gray and light-blue points are B+12 Models~1 and~2. Dashed segments connect each outlier to the nearer model median velocity within $3^\circ$ in $L_{\rm MS}$. }

    \label{fig:kinematics_results}
\end{figure}

Three of the six MagE stars within this band (Gaia DR3 2426397007431829760, 2356964085088274688, and 4993221054391991040) have $V_{\rm LSR}$ values that deviate by more than $100\,\mathrm{km\,s^{-1}}$ from the nearest local median velocity of the two B+12 models, evaluated within $3^\circ$ in $L_{\rm MS}$ (Figure~\ref{fig:kinematics_results}). In contrast, 12 of the 13 stars from \citet{Chandra2023} lie within $100\,\mathrm{km\,s^{-1}}$ of the models, with the remaining star (Gaia DR3 2661637550258741376) also being a likely kinematic outlier. We therefore exclude the three MagE outliers from our stellar MS sample, while retaining the \citet{Chandra2023} outlier for comparison. Sections~\ref{sec:correlation} and \ref{sec:orbits} present further evidence against their MS membership. Notably, the 16th--84th percentile intervals of their $L_{\rm x}$ posteriors all cross the selection boundary, highlighting the ambiguity of a selection based solely on central values. However, requiring the full $L_{\rm x}$ posterior to satisfy the selection criterion would also exclude some of our five MS candidates. The radial velocities therefore provide the additional discrimination needed to identify these outliers.

We retain five stars, two from X-shooter and three from MagE, which we refer to hereafter as the MS candidates. Together with the thirteen stars in the Chandra sample, they form the combined MS sample of 18 stars. The five MS candidates have spectro-photometric distances spanning 58--95~kpc, and their metallicities span $-1.76\leq[\mathrm{Fe/H}]\leq-1.16$. The individual measurements of the five MS candidates are listed in Table~\ref{tab:high-likelihood-candidates}. The other 46 analyzed stars, including the three velocity outliers, are listed in Table~\ref{tab:non-candidates}.
\begin{table*}
\caption{Coordinates and stellar parameters of the five MS
candidates, from both the X-shooter and MagE
samples. Metallicities are the spectroscopic values.}
\label{tab:high-likelihood-candidates}
\centering
\small
\setlength{\tabcolsep}{2.5pt}
\begin{tabular}{llrrrrrrr}
\hline\hline
Gaia DR3 source ID & Inst. & RA & Dec. & $d$ & $v_{\rm hel}$ &
$T_{\rm eff}$ & $\log g$ & $\mathrm{[Fe/H]}$ \\
 & & [deg] & [deg] & [kpc] & [$\mathrm{km\,s^{-1}}$] &
[K] & [dex] & [dex] \\
\hline
2622351759280470528
& MagE
& 335.893366 & $-7.425410$
& $95_{-18}^{+12}$
& $-103.68 \pm 0.46$
& $4298 \pm 100$
& $0.40 \pm 0.30$
& $-1.59 \pm 0.20$ \\

2441369916261698304
& X-Sh
& 359.169473 & $-7.962814$
& $90.2_{-2.0}^{+1.8}$
& $-169.58 \pm 0.04$
& $3660 \pm 100$
& $-0.35 \pm 0.30$
& $-1.75 \pm 0.20$ \\

4849744515068988544
& MagE
& 51.719178 & $-41.285058$
& $81.7_{-9.3}^{+6.5}$
& $-18.80 \pm 0.16$
& $4264 \pm 100$
& $-0.06 \pm 0.30$
& $-1.76 \pm 0.20$ \\

6503888084644067712
& X-Sh
& 341.691423 & $-57.742600$
& $57.9_{-4.2}^{+3.2}$
& $-64.92 \pm 0.07$
& $4403 \pm 100$
& $0.25 \pm 0.30$
& $-1.27 \pm 0.20$ \\

4716496865562228736
& MagE
& 19.178275 & $-60.970305$
& $57.7_{-11.4}^{+8.0}$
& $60.94 \pm 0.27$
& $4130 \pm 100$
& $0.34 \pm 0.30$
& $-1.16 \pm 0.20$ \\
\hline
\end{tabular}
\tablecomments{
X-Sh denotes X-shooter. The positions are the \gaia catalog coordinates used in the orbit
calculations. Distances are posterior medians, with the 16th and 84th
percentiles shown as asymmetric uncertainties. The quoted $T_{\rm eff}$
uncertainties include a systematic floor of $100$~K added in quadrature
to the formal uncertainties.
}
\end{table*}

Because the $L_x$ selection partly constrains proper motion, the agreement of the MS candidates with the model trends in the middle panels of Figure~\ref{fig:kinematics_results} is expected. It should therefore not be considered as independent evidence of membership, but rather a sanity check.

The spectro-photometric distances of the five MS candidates span 58--95~kpc. Their distance distribution is compared with the Cloud orbits in Figure~\ref{fig:patel_orbits}. Our MagE target selection extends to 160~kpc by construction (Section~\ref{sec:mage_selection}), so the absence of MS candidates beyond $\sim95$~kpc is not imposed by the upper distance limit of that selection. With five stars, however, the full distance extent of the stellar MS remains uncertain. The 40~kpc lower bound used for target selection also prevents this sample from testing any nearer component predicted by \citet{luccini2021} and \citet{sapna2025}.

\begin{figure*}
    \centering
    \includegraphics[height=0.355\textwidth]{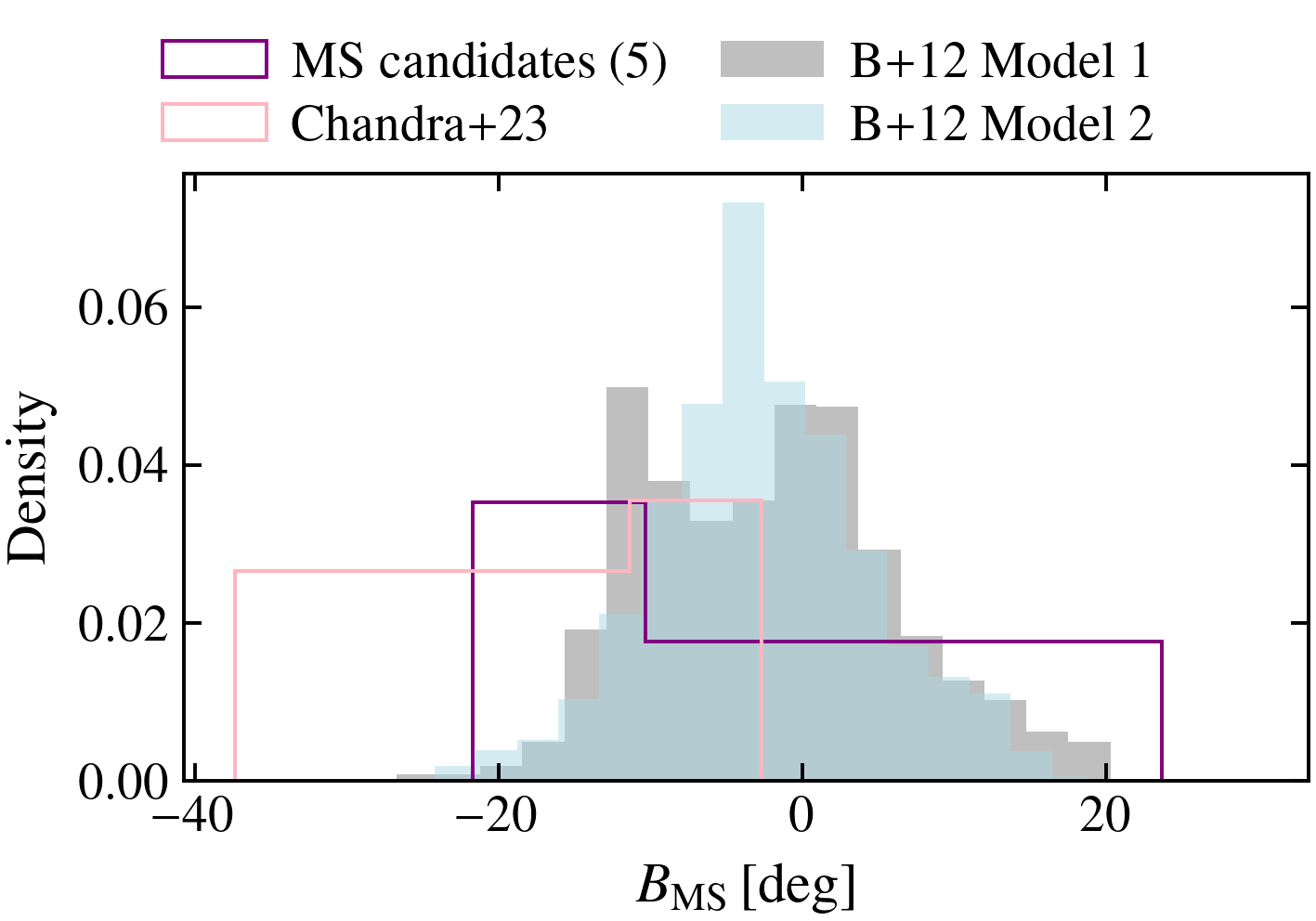}\hfill
    \includegraphics[height=0.355\textwidth]{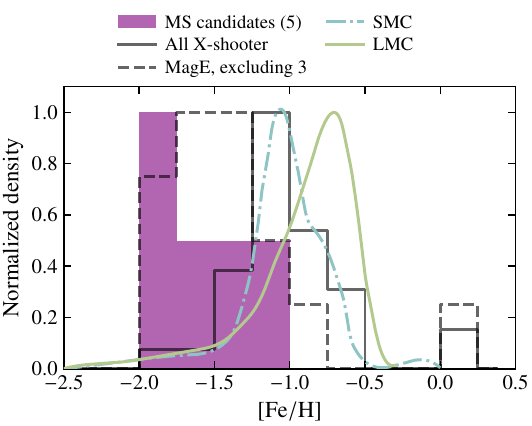}
    \caption{\textit{Left:} Magellanic-latitude distributions of the two B+12 models (gray and light blue), the Chandra sample \citep[pink line,][]{Chandra2023}, and the five MS candidates of this work (purple). The combined MS sample spans $-37^\circ<B_{\rm MS}<+24^\circ$ with median $|B_{\rm MS}|=19^\circ$, compared with $5$--$7^\circ$ for the simulations.
    \textit{Right:} Normalized metallicity distribution function of the five MS
    candidates (purple), pooling two X-shooter and three MagE stars. The black
    histograms show all 33 X-shooter targets (solid) and the 15 MagE targets
    after omitting the three radial-velocity outliers (dashed). The SMC (teal,
    dot-dashed) and LMC (sage, solid) distributions
    from \citet{munoz2023}. Every distribution is normalized to a
    peak of unity, so that samples differing by orders of magnitude in size
    remain comparable in shape.}
    \label{fig:bhist_feh}
\end{figure*}

The five MS candidates span $-22^\circ<B_{\rm MS}<+24^\circ$ (Figure~\ref{fig:kinematics_results} and the left panel of Figure~\ref{fig:bhist_feh}), with a median $|B_{\rm MS}|$ of $20^\circ$, compared with $5$--$7^\circ$ for the two B+12 models. All thirteen stars in the Chandra sample lie at negative $B_{\rm MS}$. Together, the 18 stars of the combined MS sample span $-37^\circ<B_{\rm MS}<+24^\circ$ with a median $|B_{\rm MS}|$ of $19^\circ$. The combined MS sample therefore occupies a qualitative broader band than either model predicts.

\subsection{Metallicities}
Metallicity was not used in the final candidate selection and therefore provides an independent consistency check. The \feh\ distribution of the five MS candidates is shown in the right panel of Figure~\ref{fig:bhist_feh}, alongside the LMC and SMC distributions from \citet{munoz2023}. Their median metallicity is \feh~$=-1.59$, and they span $-1.76\leq$~\feh~$\leq-1.16$. They are more metal-poor than the peak of the SMC distribution at \feh~$\simeq-1.06$ and the LMC peak at \feh~$\simeq-0.71$.

Such low metallicities are compatible with material stripped from the outer parts of the SMC. The SMC has a negative radial metallicity gradient, measured at $-0.075\pm0.011$~dex~deg$^{-1}$ over its inner $5^\circ$ from Ca\,\textsc{ii}-triplet spectroscopy of red giants \citep{Dobbie2014}. The SMC's outer population is thus expected to be more metal-poor than the central population that dominates its distribution plotted in Figure~\ref{fig:bhist_feh}. This makes the metallicities of the MS candidates consistent with an origin in the SMC outskirts \citep{munoz2023}, without imposing this metallicity restriction a priori. 

The MS candidates also fall within the metallicity range of the MW halo, so their metallicities alone do not establish membership in the Stream. Five stars are insufficient to distinguish an SMC from an LMC origin, especially as the LMC outskirts can also be more metal-poor than its center. We identified no trend between \feh\ and position along the MS. 

\subsection{Velocity gradients}\label{sec:correlation}
\begin{figure}
    \centering
    \includegraphics[width=\columnwidth]{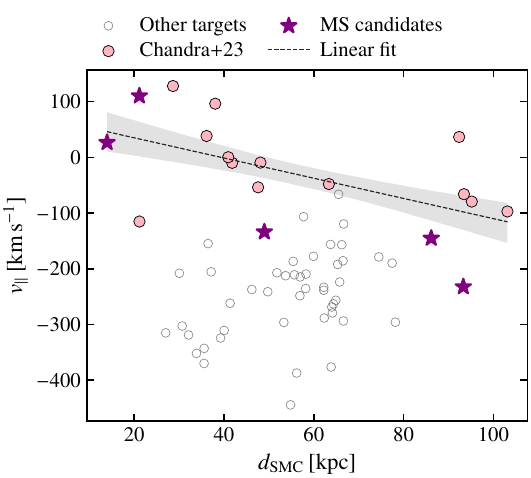}
    \caption{Velocity relative to the SMC, projected along its direction of motion, against present-day distance from the SMC. Purple stars are the five MS candidates of this work; open gray circles are the other 46 analyzed targets, including the three radial-velocity outliers. Pink circles show the Chandra sample \citep{Chandra2023}. The dashed line is a least-squares fit to the 18 stars of the combined MS sample, and the gray band is its $1\sigma$ bootstrap interval.}
    \label{fig:stream_gradient}
\end{figure}


Once stars are tidally stripped, small offsets in their orbital energies and
actions relative to the progenitor produce corresponding offsets in orbital
frequency. These differences cause the debris from the Clouds to spread into leading and
trailing tails. Consequently, within a dynamically young, unwrapped tidal
arm, stars farther along the stream from the progenitor were generally
stripped earlier, whereas recently stripped stars remain closer to the
progenitor in phase space; this ordering nevertheless has intrinsic scatter
\citep{Bovy2014,Fardal2015}. This phase-space sorting produces coherent
position--velocity gradients along tidal tails, including correlations
between the component of stellar velocity parallel to the tail and distance
from the progenitor \citep{Klimentowski2009}. Older stream segments have also
had longer to respond to gravitational perturbations, although genuine
velocity-space diffusion requires cumulative gravitational encounters
\citep{DelosSchmidt2022}. We test for a related trend across the combined MS sample using each star's present-day three-dimensional separation from the SMC and its relative velocity projected along the SMC's direction of motion. These quantities are proxies for distance and velocity along a tidal tail, so a correlation would be suggestive rather than a unique signature of SMC stripping \citep{DiazBekki2012,Zivick2019,Chandra2023}.

For this velocity-gradient calculation, we adopt the present-day SMC position and velocity from the photometric-center orbit solution of \citet{Besla2026}, described in Section~\ref{sec:orbits}. Let $(\vect{R_i}, \vect{v_i})$ be the Galactocentric position and velocity vectors of the $i^{\rm th}$ star. Let $(\vect{R_{\rm SMC}},\vect{v_{\rm SMC}})$ be the present-day Galactocentric position and velocity vectors of the SMC. The relative position and velocity vectors of the star with respect to the SMC are given by:
\begin{equation}
    \vect{R_{i/{\rm SMC}}} = \vect{R_{i}} - \vect{R_{\rm SMC}}
\end{equation}
\noindent and
\begin{equation}
    \vect{v_{i/{\rm SMC}}} = \vect{v_{i}} - \vect{v_{\rm SMC}}
\end{equation}
\noindent Let $v_{i/{\rm SMC}}^{\rm proj}$ be the component of the star's velocity relative to the SMC, projected along the SMC's direction of motion. We have:
\begin{equation}
    v_{i/{\rm SMC}}^{\rm proj} = \vect{v_{i/{\rm SMC}}} \: \cdot \: \frac{\vect{v_{\rm SMC}}}{||\vect{v_{\rm SMC}}||}
\end{equation}

Figure~\ref{fig:stream_gradient} shows $v_{i/{\rm SMC}}^{\rm proj}$ plotted against $||\vect{R_{i/{\rm SMC}}}||$. A linear fit to the 18 stars of the combined MS sample gives a gradient of $-1.82\pm0.66$~km~s$^{-1}$~kpc$^{-1}$, with Spearman $\rho=-0.61$ ($p=7.5\times10^{-3}$; Pearson $r=-0.56$, $p=0.015$). The five MS candidates alone also have a negative fitted gradient, $-3.47\pm0.95$~km~s$^{-1}$~kpc$^{-1}$, but this estimate is sensitive to individual points. This trend is compatible with position--velocity sorting of tidal debris; it does not independently establish membership. The other 46 analyzed targets are shown as open circles, including the three stars rejected by the line-of-sight velocity comparison despite passing the $L_x$ cut.

\subsection{Orbits}\label{sec:orbits}

\begin{figure}
    \centering
    \includegraphics[width=\linewidth]{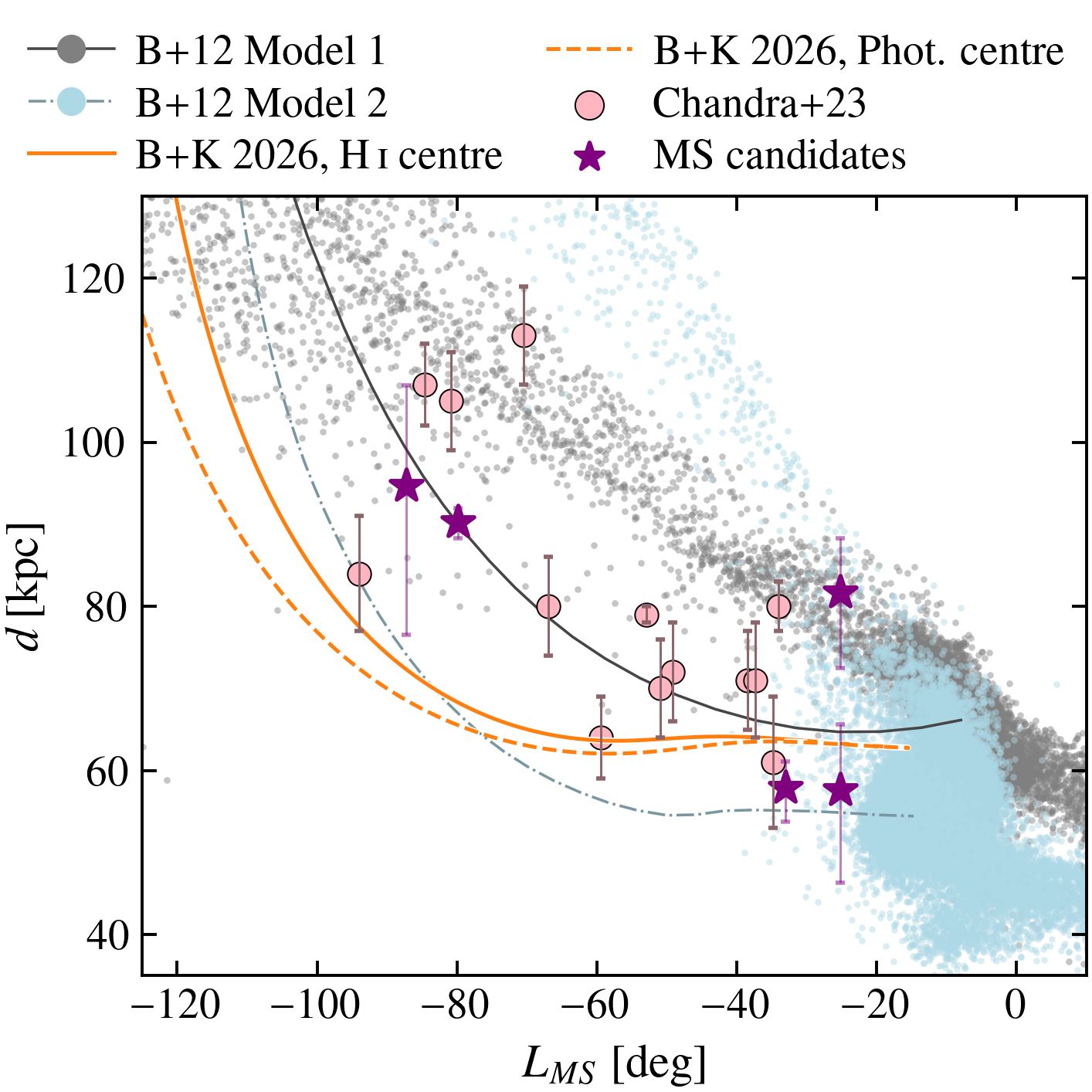}
    \caption{Magellanic longitude versus heliocentric distance. Gray and light-blue points are the stellar particles of B+12 Models~1 and~2, with their SMC center-of-mass orbits as the solid gray and dash-dotted light-blue curves. Given their origin as a tidal tail, the simulated stellar particles must reside at larger distances than the center-of-mass orbit of the progenitor SMC. The two orange curves show the SMC's trajectory over the past Gyr in the Model~C framework of B+K 2026 for a $10^{12}\,M_\odot$ MW halo: solid for the H\,\textsc{i}-center and dashed for the photometric-center orbit solutions. The Chandra sample \citep{Chandra2023} is shown as pink circles and the five MS candidates of this work are purple stars. At the MS candidates' longitudes, the B+12 simulated debris lies farther from the Sun than its corresponding SMC orbit. The H\,\textsc{i}-center and photometric-center orbits are less eccentric than the B+12 orbits, indicating that tidal debris from an SMC on the new orbits should reside at lower distances than the B+12 simulations, which adopted higher speeds for the Clouds than currently expected.
    }
    \label{fig:patel_orbits}
\end{figure}

Figure~\ref{fig:patel_orbits} plots the observed heliocentric distances of the MS candidates and the Chandra sample \citep{Chandra2023}. Despite using a completely distinct data set and selection method, we find that our MS candidates reside at distances $<$100~kpc, even though our MagE target selection extends to 160~kpc.  This further illustrates the significance of even five stars for understanding the properties of this system.

\begin{figure*}
    \centering
    \includegraphics[width=\linewidth]{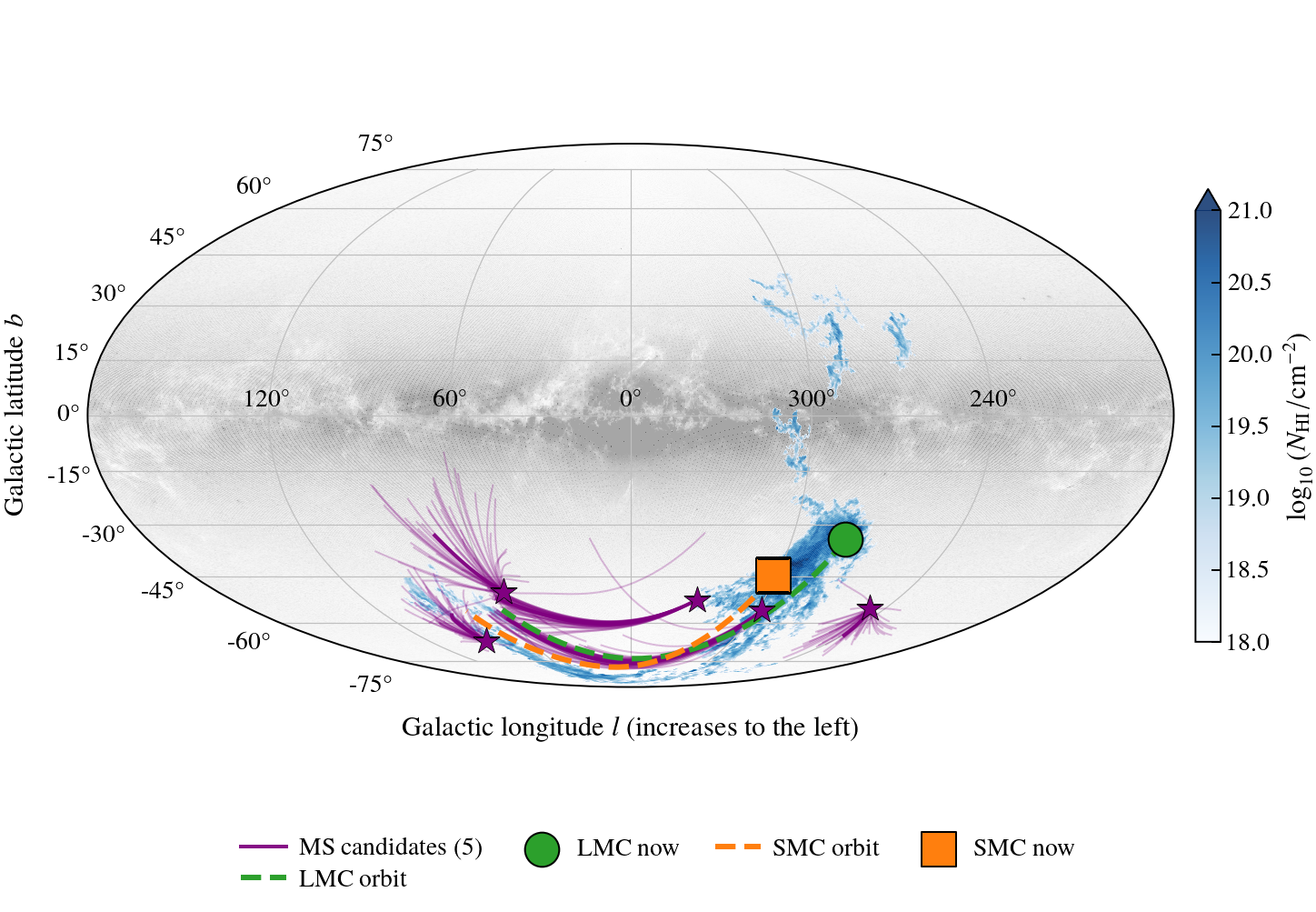}
    \caption{Mollweide projection in Galactic coordinates of the five MS candidates over the H\,\textsc{i} emission of the Magellanic system. Thick purple tracks use their measured phase-space coordinates, faint tracks sample their uncertainties, and purple stars mark their present-day positions. The three radial-velocity outliers are excluded. Dashed green and orange lines show the B+K 2026 H\,\textsc{i}-center orbits of the LMC and SMC, respectively. The H\,\textsc{i} map is from \citet{Westmeier2018}, and the grayscale background is the \gaia\ EDR3 all-sky flux map (ESA/Gaia/DPAC; A. Moitinho).}
    \label{fig:orbit_mollwide}
\end{figure*}

Figure~\ref{fig:patel_orbits} also shows the two B+12 stellar-debris models and the corresponding SMC center-of-mass orbits, together with the updated SMC trajectories from the recent review by \citet[][hereafter B+K 2026]{Besla2026}.  
The B+K 2026 orbits are computed using their Model~C moving-center-of-mass framework with the H\,\textsc{i}-center velocities \citep{Kallivayalil2013, Zivick2018} and photometric-center velocities of \cite{gaiadr3} (see their Table 2). The adopted MW model has a $10^{12}\,M_\odot$ NFW halo, disk, and bulge \citep[the MW1 model in][]{Besla2026}; the LMC model has a $1.8\times10^{11}\,M_\odot$ NFW halo and disk; and the SMC model has a $5\times10^{9}\,M_\odot$ NFW halo. These calculations include dynamical friction from the MW on both Clouds and from the LMC on the SMC (see their Appendix F).
The H\,\textsc{i}-center orbits are visualized in a Mollweide projection in Figure~\ref{fig:orbit_mollwide}; the photometric-center orbits are similar in this projection.

The distances to the B+12 simulated stellar debris are larger than those of most MS candidates. Interestingly, the B+K 2026 orbits are less eccentric than the B+12 orbits, which were based on higher speeds from the earlier measurements of \citet{Kallivayalil2006a}.
Over the longitudes occupied by our MS candidates, the stellar particles in each B+12 simulation lie farther from the Sun than the SMC orbit that generated them; the orbit is therefore not itself the predicted debris track. The stellar MS (and tidal tails in general) is expected to lie at larger distances than the SMC's orbit. We can use this dynamical fact to estimate the expected distances of MS stars removed from the SMC along the updated orbits from B+K 2026.

Around $L_{\rm MS}\simeq-80^\circ$ to $-87^\circ$, the median B+12 Model~1 debris is about 18--22~kpc farther away than its SMC orbit. The updated H\,\textsc{i}-center and photometric-center orbits lie at about 68--72 and 66--68~kpc, respectively, over the same $L_{\rm MS}$ range. Assuming a comparable debris--orbit separation based on B+12 Model~1, the expected distance of debris from an SMC following either new orbit is about 84--94~kpc.  This estimated distance range is consistent with the two candidates at 90 and 95~kpc. In contrast, the original B+12 Model~1 debris medians are about 111--117~kpc at those longitudes. We therefore anticipate that full N-body simulations using the B+K 2026 orbits will achieve better agreement with the observed distances (H. Rathore et al. 2027, in prep.).



We next attempt to confirm whether the MS candidates could have realistically originated from either the LMC or SMC by computing their trajectories in a time-evolving potential.
Using the H\,\textsc{i}-center velocities and corresponding trajectories for the MW, LMC, and SMC (described above), we constructed a time-dependent gravitational potential with \soft{AGAMA} \citep{agama}. For the test-particle orbit integrations, the LMC and SMC are represented as moving, rigid Plummer potentials with masses of $1.8\times10^{11}\,M_\odot$ and $5\times10^{9}\,M_\odot$, and scale radii of 18 and 5~kpc, respectively. 
The MW is represented using the \texttt{MilkyWayPotential2022} model of \soft{gala} \citep{gala}, which comprises spherical nuclear and bulge components, three Miyamoto--Nagai disk components \citep{MN_disk} that together approximate an exponential disk, and an NFW dark-matter halo \citep{NFW}. This model is calibrated to reproduce both the MW rotation curve measured by \citet{Eilers19} and the vertical structure inferred from the solar-neighborhood phase spiral by \citet{Darragh23}.

The B+K 2026 orbits provide the inertial phase-space coordinates of the centers of the three galaxies. We transformed the LMC and SMC orbits to a reference frame centered on the moving MW (i.e., non-inertial) by subtracting the interpolated position and velocity of the MW from each satellite at each time step. Because the stellar orbits were integrated in this accelerating Galactocentric frame, we included a spatially uniform, time-dependent fictitious acceleration associated with the motion of the MW. This term was calculated from the MW velocity trajectory as

\begin{equation}
\boldsymbol{a}_{\rm fictitous}(t)= -\frac{{\rm d}\boldsymbol{v}_{\rm MW}}{{\rm d}t}.
\end{equation}

The MW potential was held fixed at the origin of this reference frame, while the LMC and SMC potentials followed their prescribed MW-relative trajectories. The MS candidates were treated as massless test particles and therefore did not alter the trajectories or gravitational potentials of the three galaxies.

To initialize the orbit integrations, we transformed the sky positions, heliocentric distances, proper motions, and radial velocities of the MS candidates to the Galactocentric reference frame using \soft{Astropy} \citep{astropy:2013,astropy:2018,astropy:2022}. We assumed a distance to Sgr~A* of 8.122~kpc, a solar Galactocentric velocity of $(12.9,245.6,7.78)\,\mathrm{km\,s^{-1}}$, and a solar height of 20.8~pc above the Galactic mid-plane for all coordinate transformations. Figure~\ref{fig:orbit_mollwide} shows the resulting stellar orbits for the first 300 Myr on the same time grid as the LMC and SMC orbits. In projection, the five tracks cross the region occupied by the gaseous trailing Stream. This spatial overlap is consistent with, but does not prove, a Magellanic origin.

The H\,\textsc{i} background in Figure~\ref{fig:orbit_mollwide} is based on the all-sky high-velocity-cloud column-density map of \citet{Westmeier2018}, derived from the HI4PI 21-cm survey \citep{HI4PI2016}. Although Galactic disk emission is excluded by the velocity selection, the map contains non-Magellanic high-velocity complexes. We therefore isolated the Magellanic system morphologically in the Magellanic coordinate frame of \citet{Nidever2010}. After applying a threshold of $\log_{10}(N_{\rm HI}/{\rm cm^{-2}})=18$, we retained connected components larger than $35$~deg$^{2}$ with centroids within $|B_{\rm MS}|<30\degr$ and $-140\degr<L_{\rm MS}<90\degr$. This selection recovers the Clouds, Bridge, trailing Stream, and Leading Arm while excluding unrelated high-velocity H\,\textsc{i} clouds for the purpose of visualization. 

To test whether the MS candidates could have passed near either Cloud, we integrated 10,000 orbit realizations per star backward for approximately 2~Gyr, sampling their proper-motion, radial-velocity, and distance uncertainties. For these Monte Carlo integrations, we adopt the H\,\textsc{i}-center orbits of the LMC and SMC from B+K 2026 in the time-dependent potential described above. Figures~\ref{fig:posterior_orbits_1} and \ref{fig:posterior_orbits_2} show the resolved closest-approach locations relative to the LMC (blue) and SMC (orange), omitting realizations whose minimum separation occurs at an integration boundary. The probabilities quoted below are the fractions of all 10,000 realizations that enter the corresponding scale radius during the integration.

Gaia DR3 6503888084644067712 and 4716496865562228736 each have more than 99\% probability of passing inside the LMC's 18~kpc Plummer scale radius; Gaia DR3 2441369916261698304 has about 34\%. The former two have probabilities of about 14\% and 23\%, respectively, of passing inside the SMC's 5~kpc scale radius. Gaia DR3 2622351759280470528 has about 3\% probability of passing inside the LMC scale radius, while none of the realizations for Gaia DR3 4849744515068988544 enter it. These two candidates have median resolved closest approaches of roughly 28--40~kpc from the Clouds; many Gaia DR3 4849744515068988544 realizations have no resolved minimum within the integration window.

These backward integrations depend on the adopted Cloud trajectories and do not uniquely establish an origin for any one star. In particular, the rejected star Gaia DR3 4993221054391991040 also has a close LMC passage in this potential, but its line-of-sight velocity is discrepant from both local Stream-model medians and its orbit reaches the inner MW.

This analysis builds confidence that our MS candidates did indeed originate in either the LMC or SMC. 
We acknowledge that these integrations treat the Clouds as rigid potentials and therefore do not capture deformations of their dark-matter distributions during mutual interactions, which can affect backward orbit integrations \citep{foote2026}.


\section{Conclusions}
\label{sec:conclusions}

We have searched for the stellar counterpart of the MS using distant K giants selected from DELVE DR2 and VHS DR5 photometry combined with \gaia~DR3 astrometry. Our initial catalog contains 87\,763 K-giant candidates across the Southern sky, from which we selected hundreds of stars with consistent proper motions and spatial distribution as the MS. From them, we analyzed spectra of 51 targets selected by their proper motions and magnitudes to align with the B+12 predictions for the MS. Eight pass the adopted central-value $L_x$ band. Three of these have radial velocities that differ by more than $100\,\mathrm{km\,s^{-1}}$ from both local B+12 Stream-model median velocities, suggesting significantly more radial trajectories than expected for the Stream. We therefore exclude these three stars, leaving five stars that we refer to as the MS candidates. Together with the thirteen stars in the Chandra sample \citep{Chandra2023}, they form the combined MS sample of 18 stars.  

Our main results are as follows:

\begin{itemize}

\item The combined MS sample spans $-37^\circ<B_{\rm MS}<+24^\circ$, with a median $|B_{\rm MS}|=19^\circ$, compared with $5$--$7^\circ$ for the B+12 models. Our five MS candidates lie on both sides of the gaseous Stream, whereas the stars in the Chandra sample all lie at negative $B_{\rm MS}$. The observed distribution is broader than the simulated stellar-debris distributions, favoring an extended tidal debris field rather than a narrow stream.

\item Our five MS candidates have a mean heliocentric distance of $76$~kpc, a median of $82$~kpc, and a range of $58$--$95$~kpc; the combined MS sample has a mean of $80$~kpc. Over the longitudes spanned by our MS candidates, debris in the B+12 simulations lies farther from the Sun than most observed candidates. However, the updated B+K 2026 orbits, based on HST/\gaia DR3 proper motions, indicate that the SMC's orbit is less eccentric than in B+12. At $L_{\rm MS}\simeq-80^\circ$ to $-87^\circ$, the median debris in B+12 Model~1 lies 18--22~kpc farther from the Sun than its progenitor SMC orbit. Applying a comparable debris--orbit offset to the new B+K 2026 orbits suggests that an SMC tidal tail would lie at distances of 84--94~kpc at these longitudes, consistent with the observed candidate distances.
The updated SMC trajectories are therefore more promising for explaining the distances of the MS candidates, although a new debris simulation is needed for a quantitative test.

\item The five MS candidates have $-1.76\leq{\rm [Fe/H]}\leq-1.16$, with a median ${\rm [Fe/H]}=-1.59$. These values are compatible with material stripped from the metal-poor outskirts of the SMC. Our original target selection was designed to include a population resembling the Stream, without preferentially targeting stars by metallicity or selecting the metal-poor tail.  

\item The five MS candidates show a negative correlation between their present-day distance from the SMC and their relative velocity projected along the SMC's direction of motion. The fit to the combined MS sample has a slope of $-1.82\pm0.66$~km~s$^{-1}$~kpc$^{-1}$. This pattern is compatible with tidal debris, although the five-star fit is sensitive to individual points. We do not use this correlation as a separate membership criterion.

\item In the adopted time-dependent MW--LMC--SMC potential, the MS candidates have orbits consistent with an origin in either the LMC or the SMC, with pericenter passages around the Clouds within the integration window. These close approaches depend on the adopted model and do not unambiguously associate individual stars with either Cloud. Nevertheless, the analysis strengthens the evidence for a Magellanic origin.

\end{itemize}

Taken together, the kinematics and spatial distribution of the five MS candidates are consistent with a spatially extended stellar component associated with the Magellanic system. Their broad distribution across the sky challenges the narrow latitude distributions of the two B+12 models. The distance comparison favors exploring debris generated along the newer, closer Cloud trajectories derived from the latest HST/\gaia DR3 proper motions, while accounting for the offset between the progenitor orbit and stripped stars expected from tidal stripping.

Even a sample as small as five stars contributes to our understanding of the stellar MS because the final candidate selection imposes no metallicity cut and our MagE target selection extends to 160~kpc. The absence of MS candidates beyond $\sim95$~kpc is therefore not imposed by the upper distance limit of our selection. Larger samples are essential to determine the full extent of the Magellanic stellar debris and distinguish between different orbital and stripping histories. The catalog presented here contains many additional distant K-giant candidates suitable for spectroscopic follow-up, including those in regions beyond the trailing arm explored in this work.

Mapping the stellar counterpart to the MS will become increasingly feasible with upcoming large spectroscopic programs such as 4MOST \citep{Helmi2019_4MOST} and Via \citep{Via2026}, which will provide radial velocities and metallicities for large samples of distant halo stars. Improved astrometric precision in future \gaia\ data releases, particularly DR4, will reduce uncertainties in the transverse motions and orbital histories of these distant tracers. Combining this astrometry with wide-area spectroscopy will enable us to map the stellar component of the Magellanic system with substantially larger samples and place stronger constraints on the masses, past orbits, and interaction history of the Clouds.

\begin{acknowledgments}
      MC acknowledges support from the European Research Council (ERC) grant 101002511 (VEGA\_P). JC acknowledges support from the Agencia Nacional de Investigación y Desarrollo (ANID) via Proyecto Fondecyt Regular 1231345, and by ANID BASAL project FB210003.
      We acknowledge A Dictionary of Colour Combinations (ISBN 978-4-86152-247-5), which proved a valuable resource for selecting color palettes for figures, particularly for a color-blind astronomer. This work was performed using the ALICE compute resources provided by Leiden University. This work has made use of data from the European Space Agency (ESA) mission {\it Gaia} (\url{https://www.cosmos.esa.int/gaia}), processed by the {\it Gaia} Data Processing and Analysis Consortium (DPAC, \url{https://www.cosmos.esa.int/web/gaia/dpac/consortium}). Funding for the DPAC has been provided by national institutions, in particular the institutions participating in the {\it Gaia} Multilateral Agreement.

\end{acknowledgments}

\software{Astropy \citep{astropy:2013,astropy:2018,astropy:2022}, NumPy \citep{numpy}, SciPy \citep{scipy}, Matplotlib \citep{matplotlib}, pandas \citep{pandas}, AGAMA \citep{agama}, gala \citep{gala}, dynesty \citep{dynesty}, minimint \citep{minimint}, rvspecfit \citep{rvspecfit}, PypeIt \citep{Prochaska2020,PypeIt25}, iSpec \citep{BlancoCuaresma2014,BlancoCuaresma2019}, TelFit \citep{telfit}}


\bibliographystyle{aasjournalv7}
\bibliography{bib}

\FloatBarrier 

\appendix
\section{Supplementary figures and tables}

\begin{figure*}
    \centering
    \includegraphics[width=0.9\linewidth]{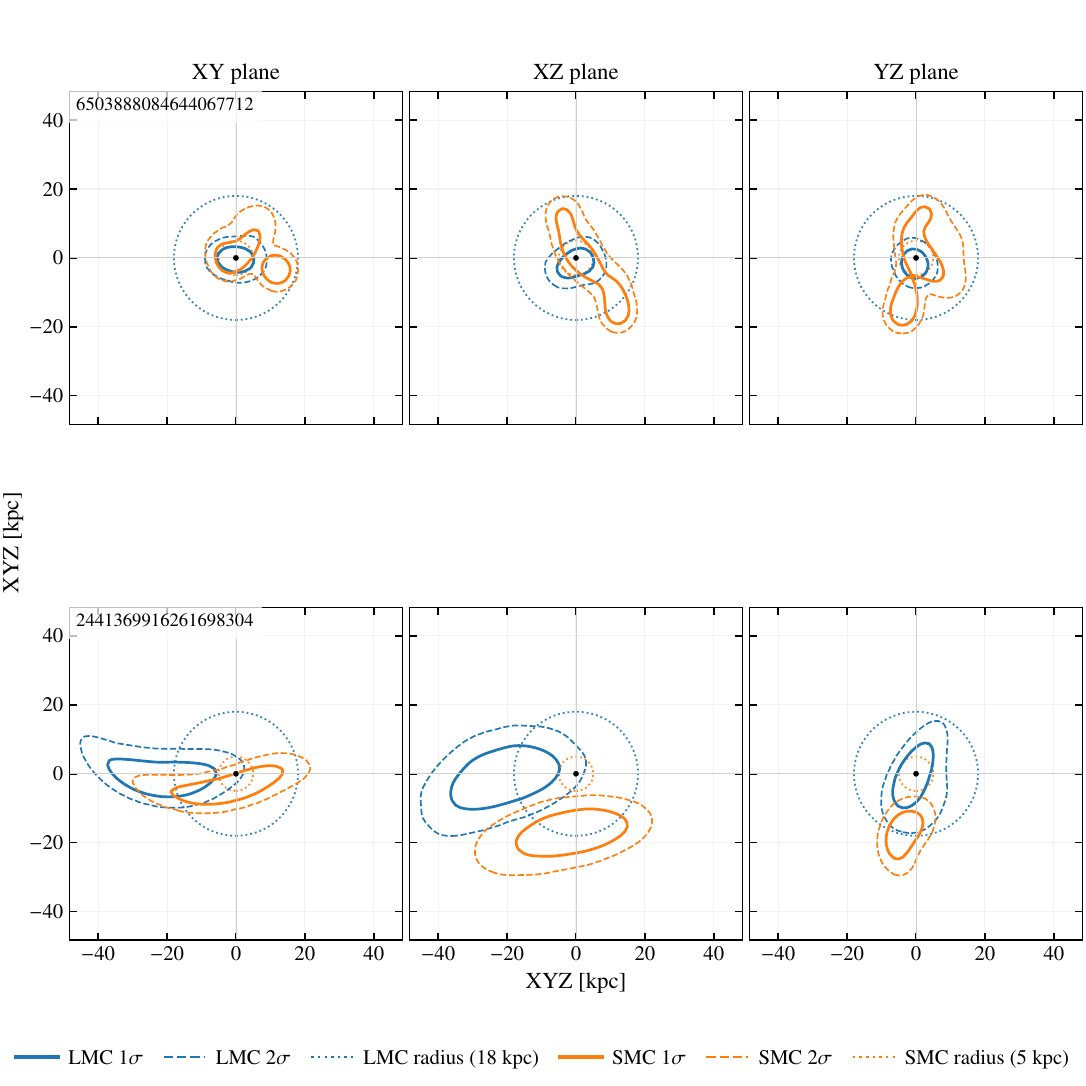}
    \caption{Posterior distributions of pericenter locations relative to the LMC center (blue) and SMC center (orange) for the two X-shooter MS candidates. The $1\sigma$ and $2\sigma$ contours are shown as solid and dashed lines, respectively. They are calculated from the resolved pericenter samples from 10,000 orbit realizations per star, with Gaussian smoothing of width 1.8 histogram bins. Realizations with a minimum separation at an integration boundary are excluded. The LMC and SMC Plummer scale radii are shown as blue and orange dotted circles, respectively \citep{Patel2020}.}
    \label{fig:posterior_orbits_1}
\end{figure*}

\begin{figure*}
    \centering
    \includegraphics[width=0.9\linewidth]{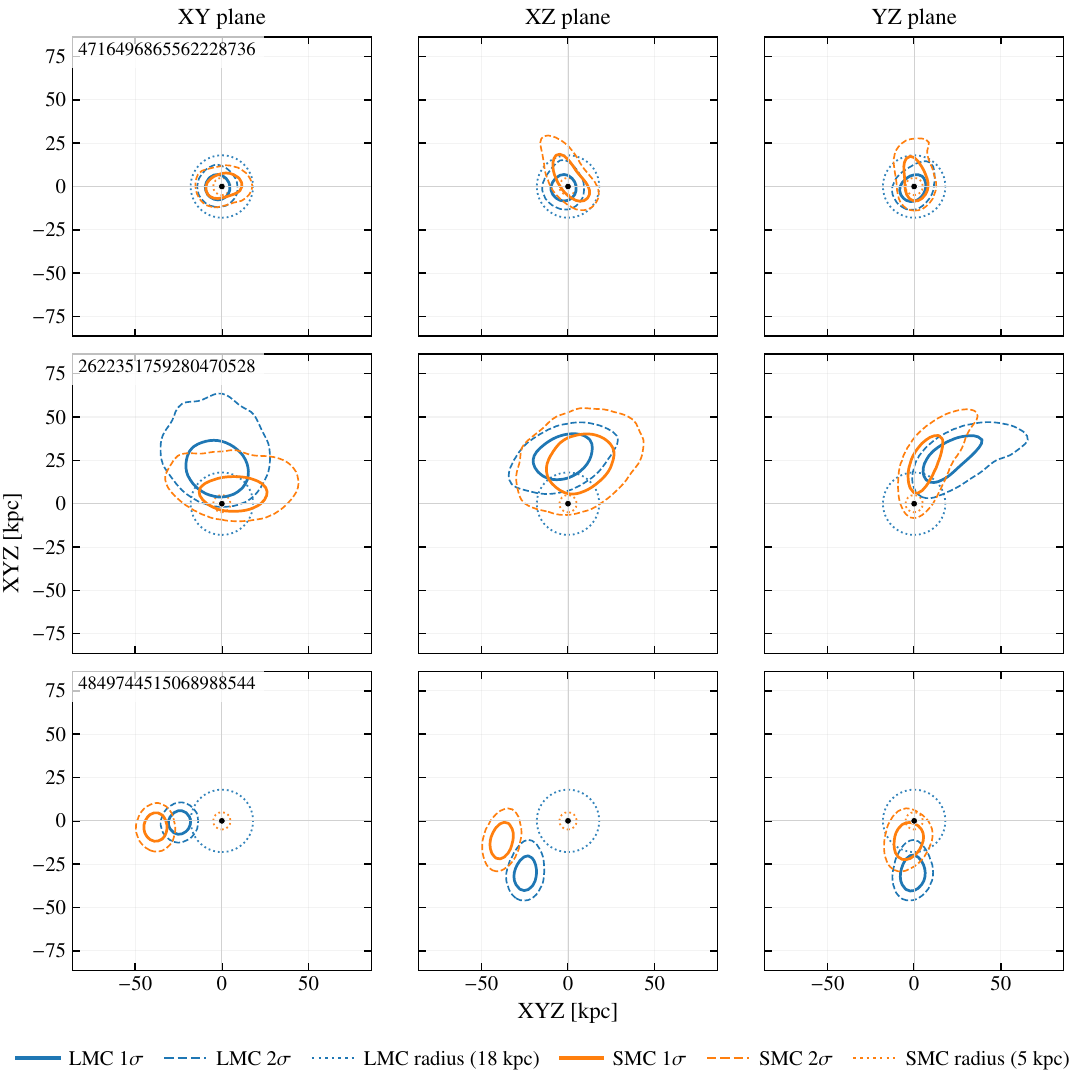}
    \caption{As Figure~\ref{fig:posterior_orbits_1}, for the three MagE MS candidates. The three radial-velocity outliers are omitted.}
    \label{fig:posterior_orbits_2}
\end{figure*}

\begin{figure*}
    \centering
    \includegraphics[width=0.8\textwidth]{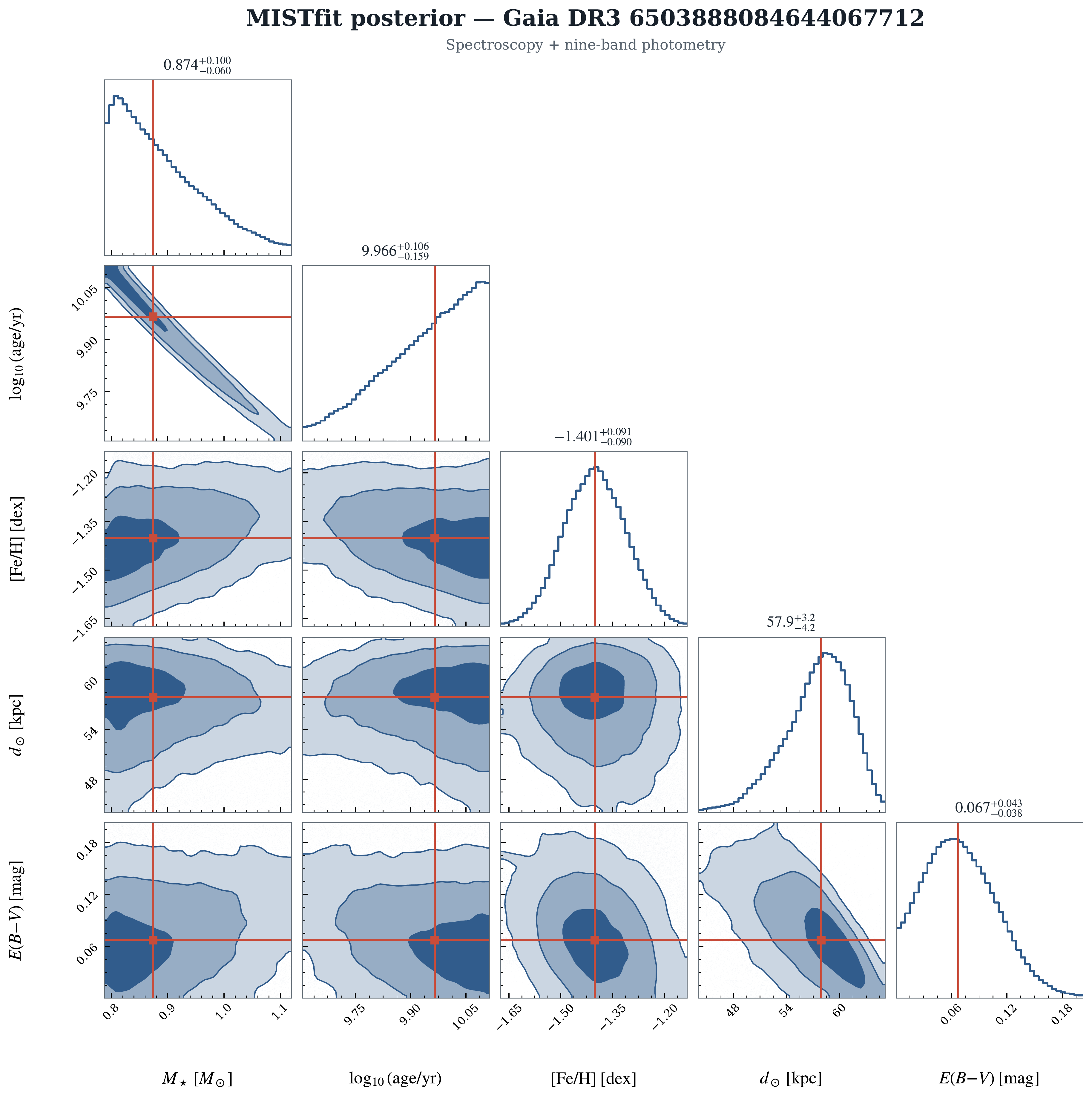}
    \caption{Example \soft{mistfit} posterior for the MS candidate
    Gaia DR3 6503888084644067712. The five parameters shown are stellar mass,
    age, metallicity, heliocentric distance and extinction; panels below the
    diagonal show the joint distributions, with contours enclosing the
    $1\sigma$, $2\sigma$ and $3\sigma$ levels, and the histograms on the
    diagonal show the marginal distributions. Red lines mark the medians. Quoted values are the medians
    with the 16th and 84th percentiles: $M_\star=0.87^{+0.10}_{-0.06}\,M_\odot$,
    $\log_{10}(\mathrm{age/yr})=9.97^{+0.11}_{-0.16}$,
    \feh~$=-1.40\pm0.09$, $d_\odot=57.9^{+3.2}_{-4.2}$~\kpc\ and
    $E(B-V)=0.067^{+0.043}_{-0.038}$. The fit combines the spectroscopic
    $T_{\rm eff}$, $\log g$ and \feh\ with nine photometric bands. The narrow
    anticorrelation between mass and age is the usual degeneracy along the
    giant branch, where a more massive star reaches the same luminosity at a
    younger age. The distance is essentially independent of it
    ($r=0.05$ against mass), which is why the spectro-photometric distance is
    well determined even though the age is not; the distance instead
    correlates with the extinction ($r=-0.70$), and this pair sets
    the achievable precision.}
    \label{fig:mistfit}
\end{figure*}

\clearpage
\begingroup
\setlength{\tabcolsep}{2.5pt}
\startlongtable
\begin{deluxetable*}{llrrrrrrr}
\tablecaption{Coordinates and stellar parameters of the spectroscopic
targets not retained as MS candidates.\label{tab:non-candidates}}
\tabletypesize{\footnotesize}
\tablewidth{0pt}
\tablehead{
\colhead{Gaia DR3 source ID} & \colhead{Inst.} & \colhead{RA} & \colhead{Dec.} & \colhead{$d$} & \colhead{$v_{\rm hel}$} &
\colhead{$T_{\rm eff}$} & \colhead{$\log g$} & \colhead{$\mathrm{[Fe/H]}$} \\
\colhead{} & \colhead{} & \colhead{[deg]} & \colhead{[deg]} & \colhead{[kpc]} & \colhead{[$\mathrm{km\,s^{-1}}$]} &
\colhead{[K]} & \colhead{[dex]} & \colhead{[dex]}
}
\startdata
2349552547787827456 & X-Sh & 11.773665 & $-22.080092$ & $83.5_{-8.4}^{+6.8}$ & $-129.66 \pm 0.09$ & $4485 \pm 100$ & $0.34 \pm 0.30$ & $-1.50 \pm 0.20$ \\
2356964085088274688 & MagE & 12.627074 & $-18.947611$ & $62.1_{-5.1}^{+4.4}$ & $45.97 \pm 0.18$ & $4157 \pm 100$ & $-0.39 \pm 0.30$ & $-1.79 \pm 0.20$ \\
2408907694443864832 & X-Sh & 352.122466 & $-13.940021$ & $19.9_{-2.3}^{+2.4}$ & $-191.60 \pm 0.14$ & $4614 \pm 100$ & $2.00 \pm 0.30$ & $-1.07 \pm 0.20$ \\
2409939891048968576\tablenotemark{a} & MagE & 345.480647 & $-15.074593$ & $50.4_{-9.9}^{+10.7}$ & $-250.48 \pm 0.50$ & $4558 \pm 100$ & $1.03 \pm 0.30$ & $-1.29 \pm 0.20$ \\
2411032809902332160 & X-Sh & 344.615379 & $-13.845066$ & $35.1_{-4.5}^{+5.2}$ & $-162.44 \pm 0.23$ & $4773 \pm 100$ & $1.63 \pm 0.30$ & $-1.00 \pm 0.20$ \\
2413362606322074240 & MagE & 347.302020 & $-10.989435$ & $35.8_{-7.7}^{+9.0}$ & $-133.25 \pm 0.51$ & $4613 \pm 107$ & $1.50 \pm 0.30$ & $-0.94 \pm 0.20$ \\
2426397007431829760 & MagE & 6.926652 & $-10.192943$ & $59.8_{-7.4}^{+5.2}$ & $17.01 \pm 0.20$ & $4249 \pm 100$ & $0.01 \pm 0.30$ & $-1.75 \pm 0.20$ \\
2444786098888864256 & X-Sh & 1.763387 & $-4.475668$ & $43.7_{-5.1}^{+5.8}$ & $-77.73 \pm 0.12$ & $4533 \pm 100$ & $1.35 \pm 0.30$ & $-1.00 \pm 0.20$ \\
2481963537665282944\tablenotemark{a} & MagE & 19.829467 & $-5.373494$ & $55.3_{-10.4}^{+9.7}$ & $-134.50 \pm 0.26$ & $4375 \pm 100$ & $0.50 \pm 0.30$ & $-1.65 \pm 0.20$ \\
2529777385642667136 & X-Sh & 12.537191 & $-2.099587$ & $46.6_{-5.1}^{+6.2}$ & $-87.31 \pm 0.09$ & $4674 \pm 100$ & $0.94 \pm 0.30$ & $-1.00 \pm 0.20$ \\
2534734362017687680 & MagE & 18.063123 & $+0.101287$ & $62.0_{-6.9}^{+5.3}$ & $-121.90 \pm 0.16$ & $4245 \pm 100$ & $-0.00 \pm 0.30$ & $-1.73 \pm 0.20$ \\
2601835941752991872 & X-Sh & 338.639362 & $-11.790160$ & $13.4_{-1.4}^{+1.7}$ & $-207.69 \pm 0.16$ & $4504 \pm 100$ & $2.50 \pm 0.30$ & $-0.50 \pm 0.20$ \\
2602183761089670784 & X-Sh & 336.252480 & $-12.133427$ & $52.5_{-5.8}^{+5.8}$ & $-139.10 \pm 0.10$ & $4557 \pm 101$ & $0.66 \pm 0.30$ & $-1.12 \pm 0.20$ \\
2606713100456070912 & MagE & 345.204876 & $-9.337777$ & $55.0_{-4.6}^{+3.7}$ & $-94.79 \pm 0.24$ & $4092 \pm 100$ & $-0.20 \pm 0.30$ & $-1.86 \pm 0.20$ \\
2607354459332524416 & X-Sh & 345.698125 & $-8.328476$ & $52.4_{-6.8}^{+7.9}$ & $-95.37 \pm 0.10$ & $4622 \pm 101$ & $0.77 \pm 0.31$ & $-1.00 \pm 0.20$ \\
2609809183465911680 & X-Sh & 340.332831 & $-7.527354$ & $55.1_{-5.1}^{+4.7}$ & $-42.10 \pm 0.09$ & $4442 \pm 100$ & $0.71 \pm 0.30$ & $-1.00 \pm 0.20$ \\
2610146806550153216 & X-Sh & 342.566724 & $-8.533369$ & $51.3_{-5.6}^{+5.9}$ & $-212.07 \pm 0.11$ & $4600 \pm 100$ & $0.88 \pm 0.30$ & $-1.16 \pm 0.20$ \\
2614779835016325120 & X-Sh & 334.517977 & $-10.598957$ & $56.0_{-5.3}^{+4.8}$ & $-242.17 \pm 0.11$ & $4600 \pm 100$ & $0.69 \pm 0.30$ & $-1.10 \pm 0.20$ \\
2615394535031185920 & X-Sh & 336.402947 & $-8.728369$ & $22.0_{-2.9}^{+3.1}$ & $-173.35 \pm 0.14$ & $4600 \pm 100$ & $1.95 \pm 0.30$ & $-1.00 \pm 0.20$ \\
2627238263832329984 & MagE & 332.790557 & $-4.179408$ & $44.5_{-7.3}^{+8.2}$ & $-293.47 \pm 0.30$ & $4500 \pm 100$ & $0.80 \pm 0.30$ & $-1.37 \pm 0.20$ \\
2627493419249681280 & MagE & 335.764306 & $-3.425821$ & $39.6_{-7.2}^{+7.7}$ & $-8.42 \pm 0.20$ & $4441 \pm 101$ & $1.13 \pm 0.30$ & $-1.41 \pm 0.20$ \\
2680349107859540096 & MagE & 331.184038 & $-0.443837$ & $40.5_{-7.5}^{+10.6}$ & $-278.19 \pm 0.58$ & $4653 \pm 100$ & $1.70 \pm 0.31$ & $-1.00 \pm 0.20$ \\
4910361621900734208\tablenotemark{a} & X-Sh & 17.417055 & $-57.531942$ & $39.4_{-5.2}^{+6.2}$ & $74.78 \pm 0.14$ & $4823 \pm 100$ & $1.40 \pm 0.30$ & $-1.00 \pm 0.20$ \\
4944973693890618880 & MagE & 38.483418 & $-47.735895$ & $29.4_{-6.1}^{+5.4}$ & $34.95 \pm 0.32$ & $4478 \pm 101$ & $2.60 \pm 0.30$ & $0.10 \pm 0.20$ \\
4977773229557246976 & MagE & 2.287267 & $-48.376980$ & $43.0_{-7.2}^{+10.3}$ & $73.64 \pm 0.28$ & $4350 \pm 100$ & $0.59 \pm 0.30$ & $-1.34 \pm 0.20$ \\
4979295743923207680\tablenotemark{a} & X-Sh & 9.507136 & $-46.389988$ & $27.7_{-3.3}^{+4.0}$ & $2.06 \pm 0.06$ & $4459 \pm 100$ & $1.30 \pm 0.30$ & $-1.00 \pm 0.20$ \\
4993221054391991040 & MagE & 9.057444 & $-41.271321$ & $70_{-14}^{+16}$ & $117.63 \pm 0.51$ & $4644 \pm 103$ & $1.00 \pm 0.30$ & $-1.27 \pm 0.20$ \\
6458074626274350080 & X-Sh & 315.346833 & $-55.198573$ & $50.9_{-8.2}^{+6.7}$ & $155.90 \pm 0.06$ & $4518 \pm 100$ & $0.50 \pm 0.30$ & $-1.18 \pm 0.20$ \\
6518522240811816320 & X-Sh & 336.689384 & $-47.029770$ & $52.3_{-5.5}^{+5.5}$ & $-39.50 \pm 0.11$ & $4376 \pm 100$ & $0.88 \pm 0.30$ & $-1.00 \pm 0.20$ \\
6518847318297206528 & X-Sh & 336.442639 & $-46.093396$ & $7.90_{-0.87}^{+1.01}$ & $6.85 \pm 0.14$ & $4723 \pm 100$ & $3.00 \pm 0.30$ & $-0.62 \pm 0.20$ \\
6532434911193453952 & MagE & 354.412162 & $-42.284482$ & $47.2_{-3.5}^{+3.3}$ & $187.63 \pm 0.15$ & $4068 \pm 100$ & $-0.50 \pm 0.30$ & $-1.84 \pm 0.20$ \\
6536701153747802368 & X-Sh & 350.799829 & $-40.746122$ & $57.2_{-9.7}^{+6.8}$ & $148.57 \pm 0.08$ & $4624 \pm 100$ & $0.50 \pm 0.30$ & $-1.36 \pm 0.20$ \\
6544916189235036672 & X-Sh & 339.468720 & $-42.522774$ & $73.1_{-6.0}^{+4.3}$ & $14.31 \pm 0.07$ & $4367 \pm 100$ & $0.19 \pm 0.30$ & $-1.15 \pm 0.20$ \\
6548490220500141440\tablenotemark{a} & X-Sh & 349.620432 & $-39.933863$ & $56.8_{-1.9}^{+1.5}$ & $-115.32 \pm 0.05$ & $3900 \pm 100$ & $-0.50 \pm 0.30$ & $-1.46 \pm 0.20$ \\
6560830997275995904\tablenotemark{a} & X-Sh & 329.940187 & $-48.464208$ & $38.9_{-5.8}^{+7.6}$ & $48.49 \pm 0.20$ & $4768 \pm 100$ & $1.50 \pm 0.30$ & $-1.16 \pm 0.20$ \\
6579489610816557184 & X-Sh & 323.029483 & $-40.589094$ & $13.2_{-1.5}^{+1.9}$ & $-84.01 \pm 0.07$ & $4238 \pm 100$ & $2.27 \pm 0.30$ & $0.10 \pm 0.20$ \\
6587583901886540288 & X-Sh & 330.281399 & $-34.954261$ & $11.8_{-1.4}^{+1.5}$ & $182.03 \pm 0.05$ & $4235 \pm 100$ & $2.27 \pm 0.30$ & $0.09 \pm 0.20$ \\
6625431222418624896 & X-Sh & 335.874706 & $-25.451754$ & $31.8_{-3.6}^{+4.1}$ & $-142.21 \pm 0.11$ & $4281 \pm 100$ & $1.59 \pm 0.30$ & $-0.59 \pm 0.20$ \\
6625450601311209344 & MagE & 335.496690 & $-25.314345$ & $54.3_{-7.7}^{+13.7}$ & $-120.88 \pm 0.18$ & $4527 \pm 100$ & $0.14 \pm 0.30$ & $-1.64 \pm 0.20$ \\
6777274572785087360 & X-Sh & 316.510296 & $-34.859239$ & $39.5_{-4.2}^{+4.8}$ & $-6.12 \pm 0.17$ & $4700 \pm 100$ & $1.32 \pm 0.30$ & $-1.32 \pm 0.20$ \\
6824145855827406976 & X-Sh & 327.965006 & $-19.981063$ & $29.4_{-3.2}^{+3.9}$ & $-223.60 \pm 0.09$ & $4533 \pm 100$ & $1.50 \pm 0.30$ & $-1.00 \pm 0.20$ \\
6887945228483051392 & X-Sh & 312.819472 & $-13.060860$ & $73.0_{-8.2}^{+7.6}$ & $-148.93 \pm 0.10$ & $4474 \pm 100$ & $0.78 \pm 0.30$ & $-0.99 \pm 0.20$ \\
6897786884048378112 & X-Sh & 320.247233 & $-7.643413$ & $15.5_{-1.9}^{+2.3}$ & $-248.45 \pm 0.13$ & $4866 \pm 100$ & $2.29 \pm 0.30$ & $-0.67 \pm 0.20$ \\
6898149830259505408\tablenotemark{a} & X-Sh & 321.745523 & $-7.241198$ & $35.9_{-4.7}^{+5.7}$ & $-151.28 \pm 0.07$ & $4892 \pm 100$ & $1.50 \pm 0.30$ & $-1.05 \pm 0.20$ \\
6908629687901160192\tablenotemark{a} & X-Sh & 315.091530 & $-8.885060$ & $69.1_{-7.8}^{+12.7}$ & $-76.43 \pm 0.14$ & $4658 \pm 100$ & $0.79 \pm 0.30$ & $-1.50 \pm 0.20$ \\
6917207283052448768 & X-Sh & 315.392877 & $-1.030620$ & $35.4_{-4.4}^{+4.9}$ & $-288.75 \pm 0.16$ & $4792 \pm 100$ & $1.56 \pm 0.30$ & $-1.00 \pm 0.20$
\enddata
\tablecomments{
X-Sh denotes X-shooter. The positions are the \gaia catalog coordinates. Distances are posterior medians, with the 16th and 84th
percentiles shown as asymmetric uncertainties. The $\log g$ and
$\mathrm{[Fe/H]}$ uncertainties include adopted systematic floors of
0.30 and 0.20 dex, respectively, added in quadrature to the formal
Hessian uncertainties.
Forty-three targets do not satisfy the central-value $L_x$ band. The other
three (Gaia DR3 2426397007431829760, 2356964085088274688, and 4993221054391991040) pass that band but are excluded from the
MS candidate sample by the additional radial-velocity check described in
Section~\ref{sec:results}.
\tablenotetext{a}{The fit has \texttt{bad\_hessian=True}, indicating
that the local covariance matrix could not be estimated reliably.
The quoted $T_{\rm eff}$, $\log g$, and $\mathrm{[Fe/H]}$ uncertainties
are therefore the adopted systematic floors alone, as described in
Section~\ref{sec:stellar_parameters}.}
}
\end{deluxetable*}
\endgroup

\end{document}